\documentclass[11pt]{article}
\pdfoutput=1
\usepackage{jheppub}
\usepackage{amssymb,amsfonts}
\usepackage{mathrsfs}
\let\savenumberline\numberline
\def\numberline#1{\savenumberline{#1.}}
\makeatletter
\renewcommand{\@seccntformat}[1]{\csname the#1\endcsname.\,\,}
\makeatother

\newcommand{\BS}{{\bf S}}

\newcommand{\CA}{{\cal A}}
\newcommand{\CB}{{\cal B}}
\newcommand{\CD}{{\cal D}}
\newcommand{\CE}{{\cal E}}
\newcommand{\CF}{{\cal F}}
\newcommand{\CG}{{\cal G}}
\newcommand{\CH}{{\cal H}}

\newcommand{\CO}{{\cal O}}
\newcommand{\CR}{{\cal R}}

\newcommand{\CU}{{\cal U}}

\newcommand{\CW}{{\cal W}}
\newcommand{\CX}{{\cal X}}
\newcommand{\CY}{{\cal Y}}
\newcommand{\CZ}{{\cal Z}}

\newcommand{\Chi}{{\mathfrak X}}

\newcommand{\SB}{{\mathscr B}}

\newcommand{\MC}{{\mathbb C}}
\newcommand{\MF}{{\mathbb F}}
\newcommand{\MR}{{\mathbb R}}
\newcommand{\MT}{{\mathbb T}}
\newcommand{\MZ}{{\mathbb Z}}
\newcommand{\p}{\partial}
\renewcommand{\tilde}[1]{\widetilde{#1}}
\renewcommand{\hat}[1]{\widehat{#1}}
\newcommand{\be}{\begin{equation}}
\newcommand{\ee}{\end{equation}}
\newcommand{\bea}{\begin{eqnarray}}
\newcommand{\eea}{\end{eqnarray}}

\newcommand{\ie}{\textit{i.e.}}

\makeatletter
\def\@fpheader{\relax}
\makeatother
\usepackage{graphicx}
\usepackage{latexsym}

\title{\ \vspace{1.5in} \\ \hbox{Tropological Sigma Models}}
\author{Emil Albrychiewicz, Kai-Isaak Ellers, Andr\'{e}s Franco Valiente and Petr Ho\v{r}ava}
\affiliation{\medskip
Berkeley Center for Theoretical Physics and Department of Physics\\
University of California, Berkeley, CA, 94720-7300, USA\medskip\\
Theoretical Physics Group, Lawrence Berkeley National Laboratory\\
Berkeley, CA 94720-8162, USA}
\abstract{With the use of mathematical techniques of tropical geometry, it was shown by Mikhalkin some twenty years ago that certain Gromov-Witten invariants associated with topological quantum field theories of pseudoholomorphic maps can be computed by going to the tropical limit of the geometries in question.  Here we examine this phenomenon from the physics perspective of topological quantum field theory in the path integral representation, beginning with the case of the topological sigma model before coupling it to topological gravity.  We identify the tropicalization of the localization equations, investigate its geometry and symmetries, and study the theory and its observables using the standard cohomological BRST methods.  We find that the worldsheet theory exhibits a nonrelativistic structure, similar to theories of the Lifshitz type.  Its path-integral formulation does not require a worldsheet complex structure; instead, it is based on a worldsheet foliation structure.}
\begin{document}
\maketitle
\section{Introduction}

The work presented in this paper is at least partially motivated by a remarkable result of Mikhalkin, established twenty years ago \cite{mikhalkin}: Certain Gromov-Witten invariants of pseudoholomorphic maps \cite{gromov,ewtsm} can be computed by first going to the tropical limit \cite{msintro,rau,mikhalkinrau,litvinov} of the maps involved, effectively reducing the problem to counting certain types of piecewise-linear maps between tropicalized manifolds.  The original proofs of this perhaps surprising and potentially computationally powerful result were rooted in abstract mathematical arguments.  Since Gromov-Witten invariants emerged from quantum field theory (more precisely, a topological sigma model \cite{ewtsm} coupled to worldsheet topological gravity \cite{lpw,vv}), we would like to understand this relation between the counting of complex and tropical maps directly in physics terms, from the first principles of quantum field theory, and in particular to find its direct path-integral realization.  In the process, we are hoping to learn new things about topological strings, worldsheet quantum field theories, and more generally about the path integral method extended to the tropical regime.

Over the past twenty years (or much longer, since there were many significant precursors of tropical mathematics long before the name ``tropical'' was coined), tropical geometry has developed into a thriving field, connected to surprisingly many areas of mathematics, physics, computer science, and more (see \cite{msintro,rau,mikhalkinrau,litvinov,viro} for an introduction to tropical geometry and its applications).  Much like real and complex geometry deal with geometric objects defined over the field of real or complex numbers $\MR$ and $\MC$, the natural objects of tropical geometry are defined over a certain semifield $\MT$, known as the ``tropical semifield'': Roughly speaking, the field operation of multiplication has been replaced in $\MT$ by addition, and the field operation of addition has been replaced by maximization.  For example, given a polynomial in real variable $x$ with real coefficients,
\be
p(x)=a_1x^{k_1}+a_2x^{k_2}+\ldots+a_nx^{k_n},
\ee
with $k_1<\ldots <k_n$ a sequence of positive integers, its tropicalized version would be given in terms of standard mathematical operations by
\be
p_\textrm{trop}(x)=\textrm{max}\left\{a_1+k_1x,a_2+k_2x,\ldots,a_n+k_nx\right\}.
\label{eetrpol}
\ee
Hence, a nonlinear structure in traditional mathematics has turned into a piece-wise linear structure in tropical mathematics.  Moreover, the coefficients of the linear terms in $x$ are integers.  Such a radical departure from more traditional geometry over fields has profound consequences:  Objects of tropical geometry are typically locally modeled by piece-wise linear polyhedra, and tropical maps between two such geometries respect this structure, including the integrality of the coefficients of the linear terms.  Calculations in the tropical setting therefore take on a very different nature compared to those in more traditional areas of geometry, often taking on a purely combinatorial form.  They bring in new and unexpected connections to other areas of mathematics and computer science, such as combinatorial polyhedral geometry, dynamical programming, and optimization algorithms.  If one can rewrite traditional problems of enumerative geometry into their equivalent tropical form, one might gain a computational and conceptual advantage, and solve problems that would otherwise be inaccessible by more traditional methods.
We would like to take this remarkable relation between classical geometry and tropical geometry and look for its direct manifestations in quantum field theory, with the hope that such a direct reformulation in the physics language can teach us new and perhaps unexpected facts about path integrals and quantization.  

In their original quantum field theory formulation, the standard Gromov-Witten invariants first appeared in topological string theory, as certain correlation functions of gauge-invariant observables in a theory whose fields are maps from the worldsheet $\Sigma$ to the target space $M$ (and referred to as the ``topological sigma model'' for short)%
\footnote{More precisely, they correspond to the observables in the topological sigma model known as the A-model.  Throughout this paper, we focus entirely on the tropicalization of the A-model, leaving the analogous consideration of the B-model (see \cite{ewmirror} for a review) outside of our scope.}
coupled to topological gravity.  Such quantum theories with topological symmetries can be given a very natural path-integral representation, using the methods of BRST quantization.  Since the maps that the path integrals of this theory localize to are (pseudo)holomorphic maps
\be
\Phi:\Sigma\to M,
\ee
the worldsheet must be equipped with a dynamical complex structure -- or, equivalently, a conformal class of dynamical metrics -- and the target space must carry an almost complex structure, which may or may not be integrable.  The underlying worldsheet quantum theory is then an example of a relativistic topological theory of the cohomological type \cite{ewcoho}, and the Gromov-Witten invariants appear among the gauge-invariant (or BRST invariant) observables of this topological theory.  How do we formulate the analog of this topological field theory framework, when the pseudoholomorphic maps are replaced by tropical maps?

This question has many ramifications:  Do the topological worldsheet theories that define the Gromov-Witten invariants have a tropical analog which would be accessible to the traditional methods of quantum field theory?  Is there a tropical analog of the path integral formulation of such field theories?  Do the techniques of the BRST quantization of gauge theories naturally extend to the tropical case?  These questions need to be answered before one can offer a purely path-integral-based proof of Mikhalkin's results.

We will see that constructive answers to these questions will take us outside of the limits of traditional worldsheet theories with relativistic invariance.  The worldsheets will no longer carry complex structures, nor will they be equipped with nondegenerate metrics.  Yet, the resulting theories are consistent, and correlation functions of their physical observables are calculable.  The worldsheet theories that we will encounter turn out to belong to the class of theories with Lifshitz-like behavior \cite{mqc,lif}, sensitive to a worldsheet foliation structure.  Clearly, explorations of any such extension of string theory beyond its traditional scope could be valuable even outside the realm of topological theories.

In fact, our second motivation for the present work originates from questions about string theory in the broader context, beyond topological, with propagating degrees of freedom.  Historically, the formulation of fundamental string theory was deeply rooted in the theory of the S-matrix in Minkowski spacetime, and therefore attached to the implicit assumption of the existence of a stable, static, eternal vacuum.  In order to study non-equilibrium systems using string theory, it would be highly desirable to relax this assumption, and formulate the string-theory analog of the Schwinger-Keldysh formalism, which is suitable for the study of systems far from equilibrium.  This formalism is based on a doubled time contour, in which the system is first evolved from the remote past into the far future, and then to the past again.  But what would string perturbation theory look like, when extended far from equilibrium?  This question was addressed using the techniques of large-$N$ dualities \cite{neq,ssk,keq}, leading to a perhaps surprising prediction:  The genus expansion of string perturbation theory, familiar from the study of equilibrium states, should undergo a refinement, whereby the string worldsheets $\Sigma$ contributing to the such should decompose into three parts:  One, $\Sigma^+$, at least roughly corresponds to the evolution forward in time, another -- $\Sigma^-$ -- corresponds to the evolution backwards in time, and finally the ``wedge region'' $\Sigma^\wedge$ connecting $\Sigma^+$ to $\Sigma^-$, corresponds to the time in the far future where the two branches of the Schwinger-Keldysh time contour meet.  It is this wedge region $\Sigma^\wedge$ whose hypothetical worldsheet description remains quite enigmatic. The large-$N$ arguments have shown that $\Sigma^\wedge$ does not represent simple Cutkosky-like cuts of the worldsheets, but it has its own topological expansion, with higher-genus $\Sigma^\wedge$ contributing to string perturbation theory.  Thus, $\Sigma^\wedge$ is at least topologically two-dimensional, exhibiting the full topological complexity of two-dimensional surfaces.  On the other hand, the structure of the ribbon diagrams in the large-$N$ analysis indicates that the \textit{geometry} of $\Sigma^\wedge$ is highly anisotropic \cite{ssk}:  The worldsheet distances in the directions connecting the boundaries with $\Sigma^+$ to the boundaries with $\Sigma^-$ are effectively scaled to zero (when the distances in the transverse worldsheet direction are held fixed).  As we will see below, very similar anisotropic features in the worldsheet path integral will emerge in the study of tropical topological theories, as well as for their non-topological cousins.

In this paper, and its sequel \cite{sequel}, we will construct -- at least in the controlled setting of a topological theory -- an example of a string theory which does not require the existence of a nondegenerate metric or a complex structure on the worldsheet.  We will see how such a construction naturally emerges when we try to make sense of path integrals for topological theories of tropical maps.%
\footnote{There are other known examples of string theories whose worldsheet path integral description involves more exotic mathematics compared to the conventional critical (super)strings; perhaps most notably, the cases of $p$-adic strings and non-Archimedean strings have been studied in considerable detail \cite{pad,padi,padii,padiii,padiv}.}
We will formulate and study the topological quantum field theories describing the matter sector, defined on a surface $\Sigma$, whose path integrals localize to the solutions of the appropriately defined tropical limit of pseudoholomorphic maps from $\Sigma$ to a target space $M$.  For the lack of a better term, we refer to such tropical topological sigma models as {\it tropological sigma models}.%
\footnote{To the uninitiated, the word ``tropological'' may appear to be a random amalgam of the words ``tropical'' and ``topological''.   However, a closer inspection reveals that the word {\it tropological} has an esteemed history going back almost two thousand years, having referred to ``the use of a Scriptural text so as to give it a moral interpretation or significance apart from its direct meaning'' \cite{tropology} (see also \cite{catholic}).   Coincidentally, this word has already been used in mathematical physics recently, in a different context for the tropical version of topological theory constructions, in \cite{tropvert}; and it also appeared previously in \cite{len}.  (We thank Yoav Len for bringing Ref.~\cite{len} to our attention.)}
Throughout this paper, we will treat the appropriate version of worldsheet gravity as a fixed, nondynamical background.  The construction of the appropriate worldsheet tropicalized topological quantum gravity that our tropological sigma models can naturally couple to will be presented in the sequel paper \cite{sequel}.  In the final parts of this paper, we will apply the lessons learned from the topological sigma-model case in a broader context, and explore some aspects of this type of theories without the restriction that the worldsheet theory be topological. 

The present paper is organized as follows.  In the rest of \S1, we provide a lightning overview of some central features of tropical geometry, focusing on those directly relevant for this paper.  Then we discuss some of the first obstacles and potential pitfalls that we are facing in an attempt to give a path-integral representation to a tropicalized sigma model, and propose our candidate for the tropical version of the localization equations.  In \S2, we study the worldsheet and target-space geometric structures associated with these tropicalized localization equations, clarifying their symmetries and highlighting the differences from the standard relativistic case.  In \S3 we construct the tropical version of the topological sigma model, using the traditional method of BRST quantization in the path integral formulation.  We specifically address some of the novelties compared to the relativistic case, in particular in the structure of antighost and auxiliary BRST multiplets, and explain how to deal with a residual gauge symmetry that did not appear in the relativistic case.  In \S4, we consider the example with the tropicalized $\MC P^1$ as the target space, solve for all its topological correlation functions of point-like observables at any genus, and confirm that the results match the correlation functions of the relativistic $\MC P^1$ topological A-model.  In \S5 we depart from the limitations of topological theories, and study the simplest bosonic sector of the tropical sigma models as a theory with propagating degrees of freedom.  We focus on the question of a proper analytic continuation of the theory to real worldsheet time.  In \S6 we conclude with some remarks about possible generalizations.  

\subsection{Tropical mathematics and Gromov-Witten invariants}

For two real numbers, $a,b\in\MR$, define a one-parameter family of two operations, labeled by $\hbar\geq 0$:  the product $\odot_\hbar$ and the addition $\oplus_\hbar$, by
\bea
e^{a\odot_\hbar b/\hbar}&=&e^{a/\hbar}\cdot e^{b/\hbar},\\
e^{a\oplus_\hbar b/\hbar}&=&e^{a/\hbar}+ e^{b/\hbar}.
\eea
For $\hbar$ real and positive, these operations equip the real numbers with the structure of a field $\MR_\hbar$, which is canonically isomorphic to the field $\MR$.  However, as $\hbar\to 0$ (and denoting $\odot_0$ and $\oplus_0$ simply by $\odot$ and $\oplus$), the rules contract to 
\bea
a\odot b&=&\lim_{\hbar\to 0} \hbar\log \left\{e^{(a+b)/\hbar}\right\}=a+b,\\
a\oplus b&=&\lim_{\hbar\to 0}\hbar\log\left\{e^{a/\hbar}+e^{b/\hbar}\right\}=\left\{
\begin{array}{l}a,\qquad \textrm{if}\ a>b;\cr b,\qquad \textrm{if}\ b>a.\cr\end{array}\right. 
\eea
Since $\oplus$ is now idempotent, $\MR_\hbar$ itself contracts to the tropical semifield, $\MR_{\hbar=0}\equiv\MT$, with $a\odot b=a+b$ and $a\oplus b={\rm max}(a,b)$.  (More accurately, $\MT=\MR\cup \{-\infty\}$, with the role of tropical unity played by 0, and the role of tropical zero played by $-\infty$.)  This procedure has been known as the Maslov (or sometimes Litvinov-Maslov) dequantization \cite{litvinov,viro,virohyper,msintro}.  

With this tool at hand, tropicalizations of complex manifolds can be generated, roughly speaking, as follows.  Consider a complex manifold $M$, of complex dimension $n$.  Choose a suitable system of complex coordinates $Z^I$ on $M$, with $I=1,\ldots , n$.  Then take the absolute values $|Z^I|$, and apply the $\hbar\to 0$ tropicalization to each of these absolute values individually.  This yields a piece-wise linear object $M_\textrm{trop}$, locally in generic points of $M_\textrm{trop}$ of real dimension $n$. Such objects are naturally parametrized by tropical coordinates, consisting of $X^I\sim\log |Z^I|$ for $I=1,\ldots, n$ in the $\hbar\to 0$ limit (and with the phases of all $Z^I$ forgotten).  The tropical coordinates $X^I$ then satisfy the tropical algebraic rules in $\MT$.  This procedure works particularly nicely for simplest and most symmetric complex manifolds which are essentially covered by one privileged coordinate system.  $\MC P^n$ would be an example, leading to tropical projective spaces $\MT P^n$, which carries the linear structure of a real $n$-dimensional polygon.  

\subsection{Reminder: Origins of tropical geometry in superstring theory and M-theory}
\label{ssremind}

Historically, one of the stronger streams that contributed to the formation of the field of tropical geometry came from string theory and M-theory.  Tropicalizations of complex curves have naturally emerged in several different corners, mostly in the context of considering dualities of various extended BPS objects.  In this \S~\ref{ssremind}, we remind the reader briefly of some of the historical context, as an additional motivation for our own treatment of the tropicalization of topological sigma models below.  Although this historical string-theory perspective reviewed in \S~\ref{ssremind} will provide heuristic insights that we will find useful for our further approach, it is not strictly necessary for the rest of this paper, and the reader not interested in the superstring/M-theory context can skip this section.

One of the first instances where it was understood how piece-wise linear supersymmetric BPS configurations of various branes ending on other branes can be lifted to a smooth holomorphic brane in M-theory was the case of Type IIA D4-branes ending on infinite NS5-branes (and possibly also with D6-branes present) \cite{ewholom}.  Consider Type IIA superstring theory on the flat Minkowski spacetime $\MR^{10}$, with coordinates $X^0,\ldots X^9$, first at very weak string coupling.  We will place parallel NS5-branes at $X^7=X^8=X^9=0$ and at various fixed values of $X^6$.  The worldvolume of the NS5-branes is thus parametrized by coordinates $X^0,X^1,\ldots,X^5$.  We connect the NS5-branes by D4-branes which are at some fixed values of $X^4$ and $X^5$, stretching along $X^6$ between two NS5-branes.  The D4-brane worldvolume is thus parametrized by $X^0,\ldots, X^3$ and $X^6$, with $X^6$ a compact interval.  This piece-wise linear network is a 1/4 BPS state in Type IIA superstring theory.

A useful vantage point can be gained by lifting this configuration of intersecting D4-branes and NS5-branes to strong coupling, described by M-theory on $\MR^{10}\times S^1$.  The radius $R$ of the extra dimension of M-theory plays the role of the Type IIA string coupling.  We will denote by $X^{10}$ the coordinate on this M-theory $S^1$.  In this picture, D4-branes and NS5-branes have the common origin, a single smooth M-theory M5-brane; depending on whether the M5-brane wraps the $S^1$ or not, it gives rise to the D4-brane or the NS5-brane at weak string coupling.  Such an M5-brane can again be viewed as located at $X^7=X^8=X^9=0$.  Its worldvolume will again be parametrized by $X^0,\ldots, X^5$.  For such an M5-brane, the condition of 1/4 BPS supersymmetry simply requires that the worldvolume should be embedded into $X^6+iX^{10}$ holomorphically, with the corresponding boundary conditions at infinity.  In particular, the configuration corresponding to the M-theory lift of our D4- NS5-system is described by the holomorphic map
\be
X^6+iX^{10}=R\sum\log (U-a_i)-R\sum\log (U-b_i);
\label{eemfive}
\ee
in terms of the complex coordinate
\be
U=X^4+iX^5
\ee
parametrizing in the directions transverse to the D4-branes. (The first sum is over all the D4-branes that end on the NS5-brane from the left, and the second sum over those D4-branes that end on the NS5-brane from the right; $a_i$ and $b_i$ are the locations of the D4-branes inside the NS5; see \cite{ewholom} for details).  In the limit of $R\to 0$, which corresponds to the weak string coupling, this smooth single holomorphic curve degenerates into the network of piece-wise linear flat branes that we started with in the Type IIA theory.  Mathematically, this relation is very reminiscent of tropicalization; indeed, the relation between $U$ and $X^6+iX^{10}$ that follows from (\ref{eemfive}) can be characterized as
\be
U\sim \exp\left\{\frac{X^6+iX^{10}}{R}\right\},
\ee
and we see the limit of zero radius $R\to 0$ of the M-theory circle corresponds to the Litvinov-Maslov dequantization limit.  The role of the Litvinov-Maslov $\hbar$ is thus played by the Type IIA string coupling, the tropical limit corresponding to the limit of zero string coupling.

Yet another prime example of tropicalization emerging from string and M-theory is given by BPS objects in Type IIB superstring theory in $\MR^{10}$.  There are two antisymmetric 2-form gauge fields $B^{(1)}$ and $B^{(2)}$ in the spacetime supergravity description of this system.  Consequently, this theory contains half-BPS strings, labeled by two co-prime integers $(p,q)$ and referred to as $(p,q)$-strings.  $p$ and $q$ are the charges under the two $B$-fields.  At weak coupling, one can interpret say the $(1,0)$-string as the fundamental string, and the $(0,1)$-string as the corresponding D-string.

Naively, one expects that the fundamental string is allowed to end on a stretched D-string; however, the rules for joining strings in this case turn out to be more sophisticated \cite{sn,sni,snii,sniii,sniv}, as they require that when a string juncture is formed, the total $p$ and $q$ charges must be separately conserved,
\be
\sum p_i=0,\qquad \sum q_i=0,
\ee
at each junction.  Thus, a (1,0)-string can join a (0,1)-string, but the third leg in this junction must be a (1,1)-string, with the appropriate orientation.  For the configuration to be 1/4 BPS supersymmetric, the strings must lie in a two-plane, and their directions in the plane are correlated with their $(p,q)$ charges.  Thus, not only must the charges be conserved at the junction, the strings must also come in under specific fixed angles.  This procedure can be iterated, with many individual junctions conforming to the same rules, and leading to BPS objects known as \textit{string networks}.  The geometry of this network consists of piecewise-linear segments connected at vertices subjected to the charge and angle conservation rules.  As it turns out, the resulting geometric object is mathematically just a tropical curve in the tropicalized $\MC P^2$ (which itself is essentially the two-dimensional flat plane, modulo the compactification locus).

The lift of this Type IIB string network configuration to M-theory is again very illuminating \cite{krogh}: All the different $(p,q)$ strings originate from a single object, the M-theory M2-brane.  Type IIB superstring theory on $\MR^{10}$ is dual to M-theory compactified on a two-torus $T^2$.  The complex structure of the $T^2$ determines the complexified Type IIB coupling constant; and we denote this modulus by
\be
\tau=\tau_1+i\tau_2.
\ee
Our coordinates $X^9$ and $X^{10}$ on the $T^2$ will satisfy the periodicity conditions
\be
(X^9,X^{10})\sim(X^9+2\pi R,X^{10})\sim(X^9+2\pi R\tau_1,X^{10}+2\pi R\tau_2),
\ee
with $R$ indicating the overall size of the $T^2$.  It is this radius that will be taken to zero to recover the Type IIB superstring limit.  The Type IIB S-duality symmetry group $SL(2,\MZ)$ acts naturally on this modulus by modular transformations, giving a geometric explanation of S-duality via M-theory.  

In this picture, the various $(p,q)$-strings simply correspond to the unique M2-brane wrapping various corresponding $(p,q)$ cycles inside the $T^2$.  On the M-theory side of this duality, 
the M2-branes which preserve the same degree of supersymmetry as the Type IIB string networks if their worldvolume is again embedded into the spacetime holomorphically.  Take the $(p,q)$-string network to lie in the $X^1,X^2$ plane.  It is then useful to pair up these two coordinates with the two compact coordinates on the $T^2$ and use the complex coordinates
\be
Z^1=X^1+iX^9,\qquad Z^2=X^2+i X^{10}.  
\ee
Further changing the variables to 
\be
U=\exp\left\{\frac{Z^1}{R}-\frac{\tau_1}{\tau_2}\frac{Z^2}{R}\right\},\qquad
V=\exp\left\{\frac{1}{\tau_2}\frac{Z^2}{R}\right\},\qquad
  \label{eemtwo}
\ee
one finds that the condition for the M2-brane to satisfy the same 1/4 BPS condition as the Type IIB $(p,q)$-string network simply reduces to the single condition of a vanishing holomorphic function
\be
f(U,V)=0,
\ee
defining a smooth holomorphic embedding of the M2-brane into the compactified spacetime of M-theory, with boundary conditions at infinity set by the choice of the $(p,q)$-strings in the corresponding Type IIB network.  Importantly, the Type IIB superstring limit results from taking $R\to 0$ in (\ref{eemtwo}), again essentially reproducing the steps of the Litvinov-Maslov dequantization.  This connection between the string networks and tropicalization was later highlighted again in \cite{raytrop}.

In the limit of the small radius of the M-theory circle, the holomorphic surface $\Sigma$ projects onto an amoeba \cite{mikhalkinamo} in the Type IIB dimensions, which in the strict zero-radius limit becomes the piece-wise linear tropical curve describing the geometry of the string network.  Thus, the limit of small radius is equivalent to the Litvinov-Maslov dequantization, the role of $\hbar$ is being played by the radius of the M-theory compactification torus.

In the descent from M-theory to Type IIB theory, the size of the $T^2$ shrinks to zero; however, one of the main lessons from string/M-theory dualities has been that it is very beneficial to keep the shape of the $T^2$ (or the M-theory $S^1$ in our Type IIA example above) as a part of the spacetime geometry, instead of dropping it altogether.  In the case of Type IIB string theory, such an extended formalism is known as \textit{F-theory} \cite{vafaf}, and it has lead to many original insights.  We will keep these string-theory lessons in mind, when we ask how we should treat the tropicalized manifolds in our proposed path integral for tropical topological sigma models.

\subsection{Tropical topological sigma models: Search for the localization equation}

In order to specify a topological field theory of the cohomological type \cite{ewcoho}, we need three basic ingredients:  A choice of the primary quantum fields, a list of symmetries that act on them (which will typically include some class of topological deformations of the fields), and the equations that we can use to gauge-fix the underlying symmetries.  The path integral for the theory is then constructed using the methods of BRST quantization, which adds to the list of our primary fields the corresponding ghosts, antighosts and auxiliaries.  The action is designed to be BRST exact modulo possible topological invariants, and the BRST symmetry can be used to show that the path integral localizes to the moduli space of the solutions of those chosen equations.

In our tropical case, none of the three choices that we need to make to specify our path integral are quite obvious.  First, let us consider the choice of fields in the path integral.  Should our fields be maps from the worldsheet to the tropicalized manifold?  And, more importantly, should the two-dimensional worldsheet on which these quantum fields live be replaced by its real-one-dimensional tropical limit?   With such tropical maps as fundamental integration variables, the hypothetical path integral would then likely have to be interpreted as a tropical integral.  While we are agnostic as to whether such notions of tropicalized quantum field theory and tropical path integrals can be made sense of, in this paper we will choose a more traditional way, and will find a conventional path-integral definition of the tropical topological sigma models in which the fundamental fields are still the same maps
\be
\Phi:\Sigma\to M
\ee
from a two-dimensional worldsheet $\Sigma$ to a $2n$ dimensional target space manifold $M$.  We interpret the tropicalization of $M$ not as a reduction to a real $n$ dimensional tropical variety $M_\textrm{trop}$, as would be common in the mathematical literature on tropical geometry; instead, we keep the same topology of the original complex $n$ dimensional manifold $M$ as a differentiable manifold of real dimension $2n$.  The tropical limit will then come from taking a singular limit of various geometric structures on $M$.

We use local real coordinates $Y^i$ on $M$, $i=1,\ldots, 2n$. A suitable choice for $Y^i$ would be, for example, the real and imaginary parts of the $n$ complex coordinates on $M$, 
\be
Z^I=x^I+iy^I,\qquad I=1,\ldots, n,
\ee
setting
\be
Y^i=\left\{\begin{array}{l}x^I,\qquad\textrm{for}\ i=2I-1,\cr
y^I,\qquad\textrm{for}\ i=2I.\cr\end{array}\right.
\label{eecompcoo}
\ee
On the worldsheet, we will use general real coordinates denoted by $\sigma^\alpha$, with $\alpha=1,2$.  Using this local parametrization, the map $\Phi$ is then represented by specifying $Y^i$ as functions of the worldsheet coordinates, $Y^i(\sigma^\alpha)$.  These will be our quantum fields on $\Sigma$.  

\begin{figure}[t!]
  \centering
    \includegraphics[width=0.6\textwidth]{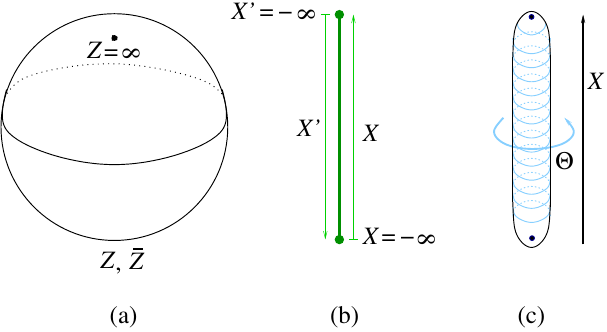}
    \caption{The tropicalization of the complex projective space $\MC P^1$. {\bf (a):} The original complex manifold $\MC P^1$.  The standard coordinate $Z\in\MC$ covers the $\MC P^1$ except at the point at infinity, and the coordinate $W=1/Z$ covers all $\MC P^1$ except the point at $Z=0$.  
{\bf (b):} The standard tropicalization $\MT P^1$, with two coordinate systems indicated: $X\in\MT$ and $X'\in\MT$. The $X$ coordinate system covers the entire $\MT P^1$ except the point at $X'=-\infty$, while the $X'$ coordinate system covers $\MT P^1$ except the point at $X=-\infty$. They overlap along their open subset $\MR$, and on this overlap the transition function must be a \textit{linear function with integer coefficients}, in this case $X'=-X$. 
{\bf (c):} The representation of $\MT P^1$ that we find suitable for our construction of the topological sigma model with $\MT P^1$ as the target space.  The underlying differential topology is the same as in $\MC P^1$, while the anisotropy between $X$ and the angular dimension parametrized by $\Theta$ results from the deformation of the geometric structures on this manifold.}
    \label{fftpone}
\end{figure}

Next, we must address the symmetries.  We will allow the theory to be gauge invariant under the standard topological deformations of $\Phi$.  In our local coordinates, these are represented by arbitrary gauge transformation functions $f^i(\sigma^\beta)$, acting simply via
\be
\delta Y^i(\sigma^\alpha)=f^i(\sigma^\alpha).
\label{eetopofs}
\ee
Our goal is to gauge-fix this topological symmetry by choosing the appropriate localization equations as gauge-fixing conditions, and applying the standard BRST formalism.

In many topological theories (such as Yang-Mills or gravity), there is also a secondary gauge symmetry, which then requires the introduction of ghost-for-ghosts in the BRST formalism.  This secondary gauge symmetry does not occur in the standard relativistic topological sigma models, and as we will see below, no such symmetries will be required in our tropical case either.

Finally, the central subtle point is the choice of the localization equations themselves.  Consider maps from a worldsheet $\Sigma$ to a target space $M$, first in the well-studied case of a relativistic topological sigma model, with $M$ which we will assume to be a complex manifold.  In this paper, we focus on the tropical version of the relativistic A-model.  In this case, the worldsheet carries a complex structure, and it is convenient use local complex coordinates $z,\bar z$ on $\Sigma$, and $n$ complex coordinates $Z^I,\bar Z^{\bar I}$ on $M$.  (A convenient choice for the $2n$ real coordinates $Y^i$ used above would then be the real and imaginary parts of $Z^I$.)

Localization equation in topological sigma models: Holomorphic maps, i.e., locally satisfying the Cauchy-Riemann equations
\be
\p_{\bar z}Z^I=0.
\label{eeholo}
\ee
In traditional algebraic geometry over the complex numbers $\MC$, one can get many examples of holomorphic maps into suitable target spaces by satisfying polynomial equations, i.e., identifying the zero loci of polynomials over $\MC$ (or over some other field $\MF$) in some ambient space (such as the projective space $\MC P^n$).  In contrast, tropical curves are not easily defined as ``tropical zeros'' of tropical polynomials over the tropical semifield: Indeed, in our example of a tropical polynomial in (\ref{eetrpol}), we see by inspection that setting it equal to the tropical zero (whose role is played by $-\infty$)
\be
p_\textrm{trop}(x)=-\infty
\ee
does not have any meaningful solutions.  Instead, tropical geometers use $p_\textrm{trop}(x)$ to define the corresponding tropical variety in a rather more indirect (and somewhat cumbersome) way:  It is the subspace consisting of all points where the polynomial is nondifferentiable; or alternatively, all points where the value of the polynomial is achieved by \textit{at least two} distinct linear terms appearing under the maximization operation.

If we wish to apply the algorithmic construction of a cohomological field theory, it would seem absolutely vital to be able to define tropical curves as objects that satisfy a certain equation.  The apparent deficiency of the tropical semifield $\MT$ to allow a definition of tropical varieties in terms of solutions of an associated tropical equation was particularly stressed by Oleg Viro \cite{virohyper,viro}, who suggested an intriguing resolution:  The tropical semifield $\MT$ (and its cousins in similar tropical constructions) should be replaced by an object that satisfies all the standard axioms of the field, except the operations in this field are sometimes multivalued.  Such generalized fields are known in the literature as ``hyperfields''.

Let us refine the discussion by considering the complex case of the appropriate hyperfield, as introduced and studied in \cite{virohyper,viro}.  Following Viro, we define the following \textit{subtropical deformation} of the field of the complex numbers.  First, introduce a map from $\MC$ to $\MC$, 
\be
S_\hbar(z)=\left\{\begin{array}{l}|z|^{1/\hbar}\displaystyle{\frac{z}{|z|}},\qquad\qquad \textrm{if}\ z\neq 0;\cr 0\qquad\qquad\qquad\qquad\textrm{if}\ z=0.\cr\end{array}\right.
\ee
Then define
\be
z\oplus_{\MC,\hbar}w=S_\hbar^{-1}\left(S_\hbar(z)+S_\hbar(w)\right).
\ee
(As pointed out by Viro \cite{viro}, a similar deformation of the complex torus $(\MC\setminus\{0\})^n$ was also originally proposed by Mikhalkin in \cite{mikhalkin}.)  If we choose to parametrize the complex numbers as
\be
z=e^{r+i\theta},
\label{eecomppara}
\ee
the $S_\hbar(z)$ map takes a particularly intuitive form in the new real variables $r,\theta$:
\be
S_\hbar(e^{r+i\theta})=e^{r/\hbar+i\theta}.
\ee
The limit of this operation as $\hbar\to 0$ then defines the complex tropical limit of the addition of complex numbers, $\oplus_{\MC,0}$.  The tropical multiplication $\odot_{\MC,0}$ will again be the standard addition.

There are two distinct ways how to interpret the $\hbar\to 0$ limit of $\oplus_{\MC,\hbar}$.  The first one follows the traditional strategy used in tropical geometry, which in the real case has lead to the semifield $\MT$, and interprets this limit as yielding univalued operation of addition, given by
\be
z_1\oplus_{\MC,0}z_2=\left\{\begin{array}{l}
z_1,\qquad\qquad\qquad \textrm{if}\ |z_1|>|z_2|;\cr
z_2,\qquad\qquad\qquad \textrm{if}\ |z_2|>|z_1|;\cr
|z_1|\displaystyle{\frac{z_1+z_2}{|z_1+z_2|}}, \ \ \quad\textrm{if}\ |z_1|=|z_2|\ \textrm{and}\ z_1\neq -z_2;\cr
0,\qquad\qquad\qquad\ \textrm{if}\ z_1=-z_2.
\end{array}
\right.
\ee
These addition rules again become more intuitive if we use the parametrization of complex numbers (\ref{eecomppara}).  In particular, the meaning of the third line is as follows:  If $r_1=r_2$ and $\theta_1\neq\theta_2+\pi (\textrm{mod}\,2\pi)$, the result of their addition is a complex number with $r=r_1$ and the phase $\theta$ which is at the midpoint between $\theta_1$ and $\theta_2$ along the shortest arc connecting them along the circle of constant $r$.  The meaning of the remaining three lines is self-explanatory.

While this single-valued operation has many good properties, it does not satisfy the axioms required of a field addition; in fact, it is not even associative.  This has lead Viro to the second interpretation of the $\hbar\to 0$ limit, in terms of multivalued operations leading to a hyperfield.  The result of such a multivalued tropical addition of two complex numbers $z_1$ and $z_2$ will now be given by specifying \textrm{a set} of values, as follows:
\be
z_1\oplus_{\MC,0}z_2=\left\{\begin{array}{l}
\{z_1\},\ \qquad\qquad\qquad\qquad \textrm{if}\ |z_1|>|z_2|;\cr
\{z_2\},\ \qquad\qquad\qquad\qquad \textrm{if}\ |z_2|>|z_1|;\cr
\{|z_1|e^{i\theta},\theta\in [\theta_1,\theta_2]\}, \ \ \quad\textrm{if}\ |z_1|=|z_2|\ \textrm{and}\ \theta_2-\theta_1<\pi;\cr
\{z\in\MC, |z|\leq|z_1|\},\qquad\ \ \textrm{if}\ z_1=-z_2.
\end{array}
\right.
\label{eevirohyper}
\ee
Thus, if two complex numbers have equal magnitude but are not opposites of each other, their addition is the shorter closed arc segment connecting them along the circle of constant magnitude; and the addition of $z$ and $-z$ is the entire closed disk of radius $|z|$.  With these rules, the addition together with the tropical multiplication satisfy the axioms of a hyperfield.  Over this hyperfield, the defining relations of a tropical variety can take the form of satisfying an equation, analogous to the polynomial equations known in classical complex geometry.

The use of hyperfields may have solved the mathematical problem of finding candidate equations whose solutions are the tropical curves, but it may have created a much greater difficulty for our formulation of the path integral:  Are we now supposed to define path integrals for quantum fields that are multivalued, and satisfy multivalued algebraic relations under addition or multiplication, and invent a hypothetical new discipline of ``quantum hyperfield theory''?  Fortunately, the answer turns out to be more prosaic, at least in the present context of the topological sigma models:  We will be able to avoid any use of hyperfields (such as Viro's hyperfield shown in \ref{eevirohyper}) altogether, and will find a path-integral realization of our theory using the conventional theory of univalued functions, and with the appropriate localization equations.  After this construction is completed, can one still find some signs of multivalued operations in our path integral?  As we will see throughout this paper, the resulting theory will exhibit a new kind of residual gauge invariance, absent in the standard relativistic topological sigma models for complex target manifolds.  We suspect that the apparent need for the multivalued operations in Viro's mathematical treatment of this problem is related to the existence of this gauge symmetry in our physical formulation.

Having been educated by the two sources -- F-theory on one hand, and the Mikhalkin-Viro treatment of the tropicalization of the complex numbers on the other -- we will keep the angular variables while performing the tropicalization of the classical geometric structures, both on the worldsheet and in the target space.  Since we found the tropicalization to be most intuitive in the parametrization of complex numbers given in (\ref{eecomppara}), our starting point will be to define similar new variables both on $\Sigma$ and on $M$,
\be
z=\exp\left\{ \frac{r}{\hbar}+i\theta\right\},\qquad Z^I=\exp\left\{\frac{X^I}{\hbar} +i\Theta^I\right\}.
\label{eetropcoo}
\ee
From now on, we will suppress the index $I$ on $Z$, since as we mentioned above, the standard tropicalization of complex manifolds is applied individually on each $Z^i$, index value by index value. In fact, we can simply consider this equation as describing a map from the punctured complex plane $z$ (with $z=0$ removed) to the punctured complex plane $Z$ (with $Z=0$ removed).  In the new variables, the holomorphicity condition becomes
\be
\p_{\bar z}Z=\frac{Z}{2\bar z}\left\{\p_r X-\p_\theta \Theta+\frac{i}{\hbar}\left(\p_\theta X+\hbar^2\p_r\Theta\right)\right\}=0.
\ee

Since we are not interested in the overall normalization while looking for a suitable localization equation, we will drop the overall normalization of the left-hand side, and keeping the leading terms in the $\hbar$ expansion both for the real and for the imaginary part.  Thus, we are led to propose
\bea
\p_r X-\p_\theta\Theta&=&0,
\label{eepxth}\\
\p_\theta X&=&0,
\label{eepx}
\eea
as our tropical localization equations.

This proposed form of the localization equations also suggests the natural use of adapted coordinates, both in the target space and on the worldsheet.  From now on, we will often use the adapted worldsheet coordinates $(r,\theta)$, and the adapted target-space coordinates
\be
Y^i=\left\{\begin{array}{l}X^I\qquad\textrm{for}\ i=2I-1,\cr
\Theta^I\qquad\textrm{for}\ i=2I,\cr\end{array}\right.
\ee
instead of the coordinates (\ref{eecompcoo}) that would have been more suitable for the standard relativistic case.  Since the tropicalization treats each $X,\Theta$ pair independently of all the others, we will often consider the simplest case of just one such pair, described by the coordinates $X,\Theta$ which we will often collectively refer to as $Y^i$, with $i=1,2$, or $Y^1=X$ and $Y^2=\Theta$.  

Let us first examine the proposed localization equations, to see if they deliver the desired solutions. Locally, the general solution of the system (\ref{eepxth}), (\ref{eepx}) is given by
\bea
X(r,\theta)&=&X_0(r),\label{eelocx}\\
\Theta(r,\theta)&=&\Theta_0(r)+\theta\p_rX_0(r),
\label{eelocth}
\eea
where $X_0(r)$ and $\Theta_0(r)$ are arbitrary (differentiable) functions of their argument.  However, globally, both $\theta$ and $\Theta$ are periodic with periodicity $2\pi$, which restricts $\p_rX_0(r)$ to be an integer, restricting the global solutions to 
\bea
X(r,\theta)&=&x_0+nr,\qquad n\in\MZ,
\label{eeglobx}\\
\Theta(r,\theta)&=&\Theta_0(r)+n\theta,
\label{eeglobth}
\eea
with $\Theta_0(r)$ still an arbitrary function.  Recall now that in the standard description of its local neighborhood, a tropical curve is described by a piece-wise linear map $X=x_0+n r$ whose slope is an integer, $n\in\MZ$.  Thus, the solutions of our localization equations describe correctly the ingredients from which tropical curves are built, when we simply apply the forgetful map $(X,\Theta)\to X$ to our pair of fields $(X,\Theta)$.  Thus, we see that at least locally in generic coordinate neighborhoods, the proposed equations indeed contain the information about the local structure of tropical maps.  We will return to the matching conditions at junctions of several local neighborhoods once we establish a covariant formulation of the localization equations, applicable to more general maps from more general surfaces $\Sigma$ to the tropical limit of the target space $M$.  While the mathematical investigations in tropical geometry are often described in the language of solely $X$, we find it useful to keep both $X$ and $\Theta$ as fundamental fields in our field theory formulation:  This will not only allow a more-or-less conventional path-integral representation of the tropical theory using ordinary quantum fields, but also lead to various clarifications, such as the origin of the quantization of $n$ (which in the reduced definition using $X$ needs to be postulated axiomatically, but which in the $(X,\Theta)$ language is simply explained as the topological winding number around $\Theta$).  

Note that there is an apparent hierarchy between the two equations in (\ref{eepxth}), (\ref{eepx}):  $\p_\theta X$ appears at one lower order in the $\hbar$ expansion than $\p_r X-\p_\theta\Theta$.  We will encounter various intriguing consequences of this hierarchy in our investigations below.

\subsection{Symmetries of the tropicalized target spaces}
\label{sssym}

Here we make a few additional clarifying remarks about the structure and symmetries of the target spaces, which will be useful during the rest of this paper.

While many complex manifolds have large continuous Lie groups of symmetries (for example those constructed as homogeneous spaces, such as $\MC P^n$), tropical manifolds exhibit only discrete symmetries \cite{msintro,mikhalkinrau}.  This is related to the fact that if they are constructed by the tropicalization limit of a complex manifold $M$, a preferred coordinate system $Z^I$ is chosen on $M$, and tropicalization is applied to each component of $Z^I$ individually, leading to a tropical coordinate system with real $X^I$ coordinates, which is adapted to the piece-wise linear and integral structure of the resulting tropical manifold.  Transition functions to another such local tropical coordinate system $\tilde X^I$ is then restricted to be piece-wise linear, with the strictly linear terms having integer coefficients $n^{IJ}$,
\be
\tilde X^I=n^{IJ}X^J+a^I.
\ee
As a result, those geometric symmetries of tropical manifolds that mix two or more $X^I$ coordinates can be at most elements of the discrete symmetry group
\be
\CG\subset GL(n,\MZ).
\label{eeglnz}
\ee
This factorization thus provides another, symmetry-based reason why we focus in the bulk of this paper on just one $X$, and its associated angle dimension $\Theta$.  Each such pair enters the description of the tropical manifold essentially independently of the others, and they are mixed together only by a subgroup of the discrete symmetry (\ref{eeglnz}).

These simple observations will have important consequences in our construction of the path integral for tropological sigma models and its symmetries.  In particular, one can question whether it is strictly necessary to formulate the theory in a fully covariant form in the target space, allowing arbitrary coordinate transformations of the $Y^i$ coordinates, or restrict only to those that respect the actual geometric symmetries of the tropical target space.  For now, we will attempt a construction which is fully covariant, and return to this issue as needed below.

Besides reducing rotational symmetries to discrete subgroups, the piece-wise linear structure of tropical geometry has another important consequence for our construction:  The reader will find that throughout this paper, the worldsheet theories we deal with are mostly free-field theories, at least when described in local adapted coordinates.  In contrast, most interesting relativistic topological sigma models correspond to targets with nonlinear geometric structures, described by highly nonlinear Lagrangians.  This is not an essential simplification on our part:  Rather, this is a feature of the tropical universe, in which nonlinear geometric structures of classical geometry have been replaced with the combinatorics of piece-wise linear structures \cite{msintro,rau,mikhalkinrau}, and are therefore amenable to a description by free quantum fields. 

\section{Geometry of the tropical limit of pseudoholomorphic maps}

We wish to be able to write our localization equations (\ref{eepxth}) and (\ref{eepx}) in a covariant form, in arbitrary coordinates.  In traditional relativistic topological sigma models, the localization equation is usually written in the covariant form as follows.  Consider maps from $\Sigma$ to $M$, at first viewed as real differential manifolds, with arbitrary real coordinates $\sigma^\alpha$ on $\Sigma$, and $Y^i$ on $M$.  First, one introduces a complex structure $\hat\varepsilon_\alpha{}^\beta$ on $\Sigma$ and an almost complex structure $\hat J_i{}^j$ on $M$.  (In this paper, we denote complex and almost complex structures $\hat\varepsilon$ and $\hat J$ with a hat, reserving the symbols $\varepsilon$ and $J$ for their tropicalized limits that will be introduced below.)  Then one demands
\be
\p_\alpha Y^i+\hat\varepsilon_\alpha{}^\beta\hat J_j{}^i\p_\beta Y^j=0.
\label{eelocrel}
\ee
We would like to rewrite our proposed localization equations in a similarly covariant form.  In order to do so, we must first understand what happens to the complex structures $\hat\varepsilon$ and $\hat J$ in the tropical limit.

\subsection{Tropicalized complex structures and Jordan structures}

Consider first the standard complex structure $\hat\varepsilon_\alpha{}^\beta$ on a relativistic $\Sigma$, which is a section of the tensor product of the tangent and cotangent bundle, $T\Sigma\otimes T^\star\Sigma$.  Starting with the complex coordinate $z\equiv u+iv$ on $\Sigma$, $\hat\varepsilon$ is simply given by
\be
\hat\varepsilon_\alpha{}^\beta d\sigma^\alpha\frac{\p}{\p\sigma^\beta}=du\frac{\p}{\p v}-dv\frac{\p}{\p u}. 
\ee
Next, we substitute our tropicalization change of variables (\ref{eetropcoo}), and express $\hat\varepsilon$ in terms of $(r,\theta)$ coordinates,
\be
\hat\varepsilon_\alpha{}^\beta d\sigma^\alpha\frac{\p}{\p\sigma^\beta}=\frac{1}{\hbar}dr\frac{\p}{\p\theta}-\hbar\, d\theta\frac{\p}{\p r}. 
\ee
As we perform the tropical contraction, $\hbar\to 0$, in order for the complex structure $\hat\varepsilon_\alpha{}^\beta$ to have a finite limit in our favorite coordinates $(r,\theta)$, we need to multiplicatively renormalize it by one power of $\hbar$, holding fixed
\be
\varepsilon_\alpha{}^\beta d\sigma^\alpha\frac{\p}{\p\sigma^\beta}=dr\frac{\p}{\p\theta}.
\ee
Similarly, we perform the tropical contraction of the target-space almost complex structure $\hat J$, leading to
\be
J_i{}^jdY^i\frac{\p}{\p Y^j}=dX\frac{\p}{\p\Theta}.
\label{eejordantar}
\ee
Note that these renormalized limits of the original complex structure do not themselves represent almost complex structures; instead, they satisfy
\be
\varepsilon^2=0,\qquad\qquad J^2=0.
\ee
As we will see in the rest of this paper, the worldsheets and target spaces in topological quantum field theories that correctly represent localization to tropical maps will carry such structures $\varepsilon$ and $J$ that square to zero (without vanishing, except perhaps at point-like singularities).  Note also that in the natural coordinates $r,\theta$ or $X^I,\Theta^I$, the resulting matrices representing $\varepsilon$ and $J$ take their Jordan normal form, with zero eigenvalues; therefore, from now on, we will refer to such structures $\varepsilon$ on the worldsheet and $J$ on the target space as \textit{Jordan structures}. 

Next, we need to write our localization equations covariantly, in terms of $\varepsilon$ and $J$.  Simply using (\ref{eelocrel}) with the tropicalized $\varepsilon$ and $J$ will not work:  These tensors are now degenerate, and imposing (\ref{eelocrel}) would lead to four independent equations, not two.  Instead, we first rewrite (\ref{eelocrel}) in a form which would be equivalent to (\ref{eelocrel}) in the relativistic case,
\be
\hat E_\alpha{}^i\equiv\hat\varepsilon_\alpha{}^\beta\p_\beta Y^i-\hat J_j{}^i\p_\alpha Y^j=0.
\label{eeloceqcovr}
\ee
We can now replace the relativistic almost complex structures in the relativistic localization equation (\ref{eeloceqcovr}) with the Jordan structures $\varepsilon$ and $J$, proposing the localization equations in the covariant form
\be
E_\alpha{}^i\equiv\varepsilon_\alpha{}^\beta\p_\beta Y^i- J_j{}^i\p_\alpha Y^j=0.
\label{eeloceqcov}
\ee
We observe that in coordinates $(r,\theta)$ and $(X,\Theta)$, these covariant equations
reduce to our desired tropical equations (\ref{eepxth}), (\ref{eepx}):
\be
E_r{}^X=\p_\theta X,\qquad E_r{}^\Theta=\p_\theta\Theta-\p_r X,\qquad E_\theta{}^X=0,\qquad
E_\theta{}^\Theta=-\p_\theta X.
\ee

\subsubsection{Self-duality and Jordan structures}

In the relativistic case, the holomorphicity condition (\ref{eeholo}) can be viewed as a self-duality relation in two dimensions.  Note that, similarly, the expression $E_\alpha{}^i$ that we intend to use to define our localization equation also satisfies a condition that reduces its number of independent components from four to two.  In adapted coordinates, we find
\be
E_\theta{}^X=0,\qquad E_\theta{}^\Theta=-E_r{}^X,
\ee
with $E_r{}^\Theta$ unconstrained.  These relations are the Jordan-structure analogs of the self-duality equations.  Can they be written in a covariant form?  The answer is yes, they are equivalent to
\be
\varepsilon_\alpha{}^\beta E_\beta{}^i=-J_k{}^i E_\alpha{}^k.
\label{eetrself}
\ee
Similarly, as in the relativistic case with the complex structure, one could define an anti-self-duality condition for generic sections $\CE_\alpha{}^i$ of $T^*\Sigma\otimes \Phi^\star(TM)$ by 
\be
\varepsilon_\alpha{}^\beta\CE_\beta{}^i=J_k{}^i \CE_\alpha{}^k.
\ee
While there are some similarities with the standard notion of self-duality on $\Sigma$ with a complex structure, there are also significant differences.  For example, while in the case of the complex structure the self-duality and anti-self-duality conditions decompose the corresponding tensors into the sum of their self-dual and anti-self-dual parts, in the case of the Jordan structure such a decomposition does not occur.  To understand this phenomenon better, let us first investigate more carefully the structures induced on a tensor algebra by the existence of a Jordan structure on a vector space.

\subsubsection{Jordan structures on vector spaces}

Consider a two-dimensional vector space $V$, such as the tangent space $T_p\Sigma$ of the two-dimensional worldsheet surface $\Sigma$ at some general point $p$.  We define the \textit{Jordan structure} on $V$ as a nonzero element $\varepsilon$ of $V^\ast\otimes V$ which satisfies, when interpreted as an endomorphism of $V$, the condition
\be
\varepsilon^2=0.
\ee
A choice of a Jordan structure $\varepsilon$ induces various unique structures on the tensor algebra associated with $V$.

It is natural to represent such a Jordan structure in an adapted basis, in which it takes the following canonical form, 
\be
\varepsilon_\alpha{}^\beta=\begin{pmatrix}0&1\\ 0&0\end{pmatrix}.
\label{eecane}
\ee
Vectors $v\in V$ annihilated by $\varepsilon$ form a one-dimensional vector subspace, $F^1V\subset V$.  It is natural to view this subspace as defining a natural filtration structure on $V$, consisting of subspaces
\be
F^0V\equiv 0\subset F^1V\subset F^2V\equiv V.
\ee

On the dual vector space $V^\ast$, the Jordan structure also induces a unique filtration.  Define $F_1V^*$ to consist of all the elements $\omega\in V^\ast$ which annihilate the elements $v\in F^1V$,
\be
\omega(v)=0.
\ee
This naturally defines a filtration
\be
F_0V^\ast\equiv 0\subset F_1V^\ast\subset F_2V^\ast\equiv V^\ast.
\ee
Of course, these filtrations of $V$ and $V^\ast$ naturally induce the corresponding filtrations of the full tensor algebra over $V$.  Note that the vector space naturally dual to $F^1V$ is \textit{not} a subspace of $V^\ast$; instead, it is the coset space $V^\ast/F_1V^\ast$.

This filtration structure on the tensor algebra is to be contrasted with the standard Hodge decomposition of the (complexified) tangent space of a $\Sigma$ that carries a complex structure, and the subsequent Dolbeault decomposition of the tensor algebra over such vector spaces with a non-degenerate complex structure.  In our case, $\Sigma$ will not be equipped with a complex structure, and the standard language of Riemann surfaces so familiar from relativistic (topological) sigma models and string theory would be unnatural.  Instead, we will develop an understanding of the natural structures and symmetries that are induced on various geometrical features of $\Sigma$ simply by postulating the existence of a Jordan structure.

\subsection{Worldsheets with Jordan structures}

Our intention is to construct topological sigma models, and later on topological strings, without having to introduce conventional complex structures (or conventional conformal structures) on the worldsheet, replacing them with the Jordan structures instead.  In order to prepare for this construction, we need to examine the consequences of the existence of a Jordan structure on $\Sigma$ and its symmetries, at first locally and then globally.

\begin{figure}[t!]
  \centering
    \includegraphics[width=0.7\textwidth]{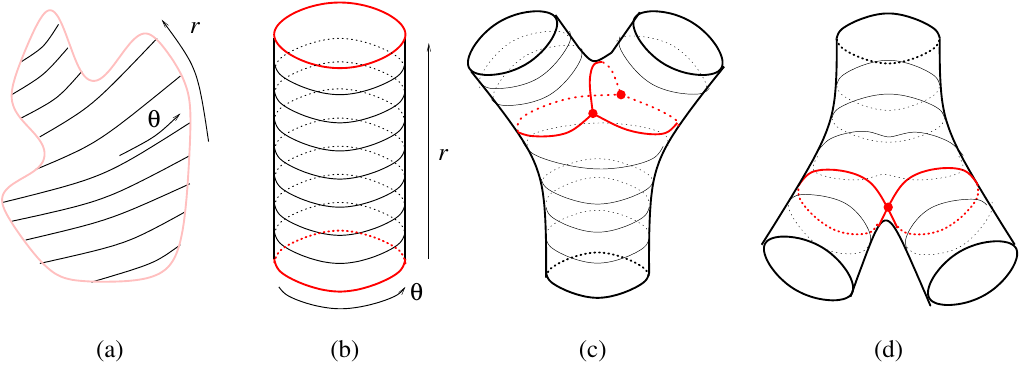}
    \caption{Examples of worldsheets with Jordan structures.  \textbf{(a):} A local open simply-connected neighborhood with a non-singular foliation (see \S\ref{sslocjor} and \S\ref{ssjsym}).  \textbf{(b):} The open cylinder with the standard non-singular foliation, referred to as ``the sleeve'' in the body of this paper (see \S\ref{sssleve}).  We expect that at the boundaries of the cylinder, the sleeve is connected to other sleeves via a singular leaf of the foliation. \textbf{(c,d):} Two examples of Jordan structures with a singular leaf, connecting several sleeves to form pieces of higher-genus geometries of $\Sigma$.  The singular leaf and therefore the Jordan structure in \textbf{(c)} is more generic than the one in \textbf{(d)}, since the singular leaf in \textbf{(d)} comes from that in \textbf{(c)} by collapsing one of the three segments of the singular leaf to a point.}
\label{ffwsjordan}
\end{figure}

\subsubsection{Jordan structures in a local neighborhood on $\Sigma$}
\label{sslocjor}

Consider an open, simply connected neighborhood $\CU$ of a generic point $p$ on the worldsheet, with a non-degenerate Jordan structure defined on $\CU$.%
\footnote{``Non-degenerate'' here means that $\varepsilon$ is chosen smoothly and is nonzero everywhere in $\CU$.  When we later consider compact $\Sigma$ of arbitrary genus, we will be naturally led for global topological reasons to allow Jordan structures over compact worldsheets $\Sigma$ that require point-like ``degenerations'' of $\varepsilon$, i.e., $\varepsilon$'s that exhibit isolated zeros or poles at a finite number of points in $\Sigma$.  Here we first focus on the non-degenerate case in simply connected open neighborhood $\CU$.}
The filtration on $T_p\Sigma$ extends smoothly over $\CU$, and defines a distribution (in the sense of vector-space subspaces) of the tangent bundle to $\Sigma$ over $\CU$.  Since this distribution is one-dimensional, it is automatically integrable, and therefore induces a natural unique foliation structure of $\CU$.  The leaves of this foliation are open intervals.  It also makes sense to require that the leaves are compact, given the fact that our construction originated from the tropical limit of a local complex coordinate system, in which the leaves of the foliation were naturally compact and parametrized by the periodic coordinate $\theta$.  Ultimately, which types of foliation should be allowed globally is an important dynamical question, which however belongs to the discussion of dynamical gravity.  In this paper, we will simply take the fixed foliation of $\Sigma$ as fixed, and given a priori, without yet making gravity dynamical.  

\subsubsection{Symmetries of the Jordan structure in a local neighborhood on $\Sigma$}
\label{ssjsym}

Consider again an open, simply connected neighborhood $\CU$ of a generic point $p$ on the worldsheet, with a coordinate system $r,\theta$ adapted to the preferred foliation induced by the Jordan structure $\varepsilon$.  (In this adapted coordinate system, $\varepsilon$ takes the canonical form (\ref{eecane}).)

The group of symmetries that preserve the choice of $\varepsilon$ is given by
\bea
\tilde r&=&\tilde r (r),\\
\tilde\theta&=&\tilde\theta_0(r)+\theta\,\p_r\tilde r(r),
\eea
with $\tilde\theta_0(r)$ an arbitrary differentiable function of its argument, and the condition $\p\tilde r/\p r\neq 0$ separating the group into two disconnected components, labeled by the sign of $\p\tilde r/\p r$.  Both connected components of the symmetry group preserve the orientation of $\Sigma$.  We will focus on the Lie algebra generators of the infinitesimal symmetries,
\bea
\delta r&=&f(r),
\label{eeinfconfr}\\
\delta\theta&=&F(r)+\theta\,\p_r f(r).
\label{eeinfconft}
\eea
Thus, the local symmetries of the Jordan structure are generated by the infinite-dimensional Lie algebra whose elements are parametrized by two real, arbitrary projectable differentiable functions $f(r)$ and $F(r)$ on the foliation.  The Lie algebra takes the natural form of a semi-direct sum, with a very clear geometric interpretation:  While $f(r)$ generates all diffeomorphisms of $r$, the infinitesimal transformations of $\theta$ are $r$-dependent affine transformations along the leaves of the foliation, with the coefficient of the term linear in $\theta$ being uniquely determined by the infinitesimal reparametrization of $r$ and indicating how the infinitesimal diffeomorphisms of $r$ act on the Abelian subalgebra of the $r$-dependent translations of $\theta$ generated by $F(r)$.%
\footnote{The reader should be warned that such a Lie algebra and its corresponding Lie group are only formally defined:  When the range of $r$ is $\MR$, there are many inequivalent definitions that make the group at least a Fr\'echet space.  They depend on the precise conditions imposed on the allowed functions $f(r)$ on $\MR$, see for example \cite{mimu}.  In this paper, we will not attempt to identify which one of these mathematically more precisely defined groups should be the best candidate for the symmetries of our path integral.}

Now we are ready to discuss the transformation properties of our proposed localization equations (\ref{eeloceqcov}).  That equation is already in a manifestly covariant form under all worldsheet diffeomorphisms generated by any $\xi^\alpha(\sigma^\beta)$, if (as we are assuming) $\varepsilon_\alpha{}^\beta$ transforms as a tensor of rank $(1,1)$, and $X$ and $\Theta$ transform as a scalar.  As a consequence, the localization equations will take the same special form (\ref{eepxth}) and (\ref{eepx}) in all coordinate systems in which the Jordan structure takes the canonical form (\ref{eecane}), if we postulate that $X$ and $\Theta$ transform under (\ref{eeinfconfr}), (\ref{eeinfconft}) as scalars,
\bea
\delta X&=&f(r)\p_r X+\left(\vphantom{\frac{1}{2}}F(r)+\theta\p_r f(r)\right)\p_\theta X,\\
\delta\Theta&=&f(r)\p_r \Theta+\left(\vphantom{\frac{1}{2}}F(r)+\theta\p_r f(r)\right)\p_\theta \Theta.
\label{eeinfconfsc}
\eea
From now on, we will assume such transformation properties for any pair $X,\Theta$ representing a tropicalized (complex) target-space dimension.

The infinite-dimensional symmetry algebra of the Jordan structure has some interesting finite-dimensional subalgebras.  First of all, the rigid translations along $r$ and $\theta$ together with the nonrelativistic boosts
\be
\delta \theta=\lambda r,\qquad \delta r=0,
\ee
form a three-dimensional nilpotent Lie algebra, well-known in the literature as the Heisenberg-Weyl algebra (generated by a canonical $p$ and $q$ pair).  Note that the role of the central element $[q,p]=i\hbar$ is played by the generator of translations in $\theta$, along the leaves of the foliation.

In fact, the algebra of all affine symmetries of the Jordan structure (i.e., symmetries acting up to linearly on $r$ and $\theta$) is not three-dimensional, but four-dimensional; it is generated by the generators of the Heisenberg-Weyl algebra, together with the isotropic constant rescalings of $r$ and $\theta$.  All four-dimensional real Lie algebras were fully classified in 1963 by Mubarakzyanov \cite{muba,low}; our algebra of all affine symmetries of $\varepsilon_\alpha{}^\beta$ appears on Mubarakzyanov's list as $A_{4.8}$.  It is an indecomposable, solvable Lie algebra.  In fact, this algebra belongs to a family, parametrized by a real parameter $-1\leq b\leq 1$, with our case corresponding to $b=0$.  In our physical interpretation, the deformations away from $b=0$ would correspond to the possibility of making the constant rescaling anisotropic, with a dynamical critical exponent
\be
z=1/(1+b).
\ee
This means that in such a deformed theory, one would be assigning an anomalous, nonzero scaling dimension to $\varepsilon$.  From now on, we will focus on the $b=0$ case.  Note that the central element of the Heisenberg-Weyl algebra is no longer central in $A_{4.8}$.

\subsubsection{Symmetries of the Jordan structure on a cylinder}
\label{sssleve}

We will be particularly interested in the foliations by compact $S^1$ leaves, which take locally the form of a $I\times S^1$, with $I$ an open interval.  We will refer to such a portion of $\Sigma$ that respects this $I\times S^1$ foliation structure as a ``sleeve''.

When we take into account the global topology of the sleeve, it is natural to restrict our class of adapted coordinates such that the $\theta$ coordinate (along the compact $S^1$ leaves of the foliation) is always periodic with periodicity $2\pi$.  This additional normalization condition will restrict the infinitesimal symmetries of the Jordan structure just to 
\bea
\delta r&=&r_0,\\
\delta\theta&=&F(r),
\eea
the Lie algebra generated by global translations of $r$ (if the sleeve is infinite and the range of $r$ is $\MR$), together with $r$-dependent rotations of $\Theta$.

One can also ask what happens if the requirement that the map from the sleeve to itself be an isomorphism is relaxed, and one simply requires it to be an endomorphism, not necessarily one-to-one; the endomorphisms of the sleeve that preserve the Jordan structure must respect the periodicity of $\theta$, and are therefore given by
\bea
\tilde r&=&nr+r_0,\qquad n\in\MZ,\\
\tilde\theta&=&\tilde\theta_0(r)+n\theta.
\eea
This is again matching the intuitive expectations from the mathematicians' treatment of tropical manifolds:  After applying the forgetful operation of dropping $\theta$, morphisms between tropical manifolds are realized by piece-wise linear maps in $r$, with the coefficients of the terms strictly linear in $r$ restricted to be integers.

Note that since we have restricted our coordinate systems on the sleeve to respect the periodicity of $\theta$ with the period of $2\pi$, this now allows us to associate an invariant length to the sleeve along the $r$ direction, without having to introduce any additional metric.  This length is simply defined as $\int dr$, between the two endpoints of the sleeve.  Thus, our foliation carries a natural measure along the tropical direction; this construction is closely reminiscent of Thurston's definition of ``measured foliations'' \cite{thurston} of two-dimensional surfaces, which plays a central role in Thurston's theory of the geometry of two- and three-manifolds.

\subsection{Anisotropic conformal symmetries}

Simply by asking what are the symmetries of a Jordan structure on $\Sigma$, we have found infinite-dimensional symmetry algebras quite reminiscent of nonrelativistic conformal symmetries in two dimensions.  In order to clarify in what sense these symmetries should be interpreted in our tropical context as ``conformal'', we will now relate the symmetries of the Jordan structure to the scaling symmetries of the appropriately defined degenerate metric structures on $\Sigma$.

\subsubsection{Tropicalization of the metric}
\label{sstropmet}

Just as we studied what happens with the standard complex structure on $\Sigma$ under the tropicalization limit, one could play the same game with a nondegenerate worldsheet
metric $\hat g_{\alpha\beta}$.  (Again, we denote nondegenerate Riemannian metrics by symbols with the hat, reserving symbols such as $g_{\alpha\beta}$ for their degenerate tropicalized limit, which we are about to introduce.)  Let us begin with the simplest case, of the flat metric with components $\hat g_{\alpha\beta}=\delta_{\alpha\beta}$ on $\Sigma$,
\be
\hat g_{\alpha\beta}d\sigma^\alpha d\sigma^\beta =du^2+dv^2,
\ee
and switch again to the tropical coordinates $r$ and $\theta$,
\be
\hat g_{\alpha\beta}d\sigma^\alpha d\sigma^\beta=\frac{1}{\hbar^2}dr^2+d\theta^2.
\ee
In order to keep the resulting object finite as $\hbar\to 0$, we must renormalize $\hat g_{\alpha\beta}$ by a multiplicative $\hbar^2$ factor, obtaining
\be
g_{\alpha\beta}d\sigma^\alpha d\sigma^\beta=dr^2.
\label{eetropmet}
\ee
This object is still a symmetric tensor, but it is no longer nondegenerate -- instead, it is of rank one on $\Sigma$, and it is the natural tropicalized candidate for replacing the relativistic metric in our nonrelativistic environment of tropicalized worldsheets.  
What are the natural conformal symmetries associated with such a degenerate metric, and are they related to the symmetries of the Jordan structure?  The appropriate notion of (anisotropic) conformal symmetry on spaces with foliations was defined in {\cite{aci}).  Given a (degenerate or nondegenerate) metric $g$ its conformal symmetry algebra simply corresponds to those allowed diffeomorphisms that map $g$ to itself up to a (possibly anisotropic) Weyl transformations.  In our case, we do not need anisotropic scaling with a dynamical exponent $z\neq 1$, and will simply restrict our attention to isotropic Weyl transformations.  The infinitesimal diffeomorphisms $\xi^\alpha$ that map (\ref{eetropmet}) to itself up to an infinitesimal Weyl rescaling by $\omega(\sigma^\alpha)$ satisfy
\be
g_{\alpha\gamma}\p_\beta\xi^\gamma+g_{\beta\gamma}\p_\alpha\xi^\gamma=\omega(\sigma^\gamma)g_{\alpha\beta}.
\ee
In our simplest degenerate case (\ref{eetropmet}), these equations reduce to
\be
\p_\theta\xi^r=0,\qquad 2\p_r\xi^r=\omega(r,\theta).
\ee
The solutions give the algebra isomorphic to the algebra of \textit{all} foliation-preserving diffeomorphisms of $\Sigma$,
\be
\xi^r=f(r),\qquad\xi^\theta=F(r,\theta),\qquad\omega=2\p_r f.
\ee
a symmetry strictly larger than the symmetry of the Jordan structure defining the foliation.

While it is somewhat intriguing that the algebra of all foliation-preserving diffeomorphisms can be interpreted as the conformal algebra associated with a degenerate metric, and isotropic Weyl transformations, this is not the end of the story.  The original metric $\hat g_{\alpha\beta}$ had its inverse metric, which we will denote by $\hat h^{\alpha\beta}$.  Writing the inverse metric in the tropical coordinates,
\be
\hat h^{\alpha\beta}\frac{\p}{\p\sigma^\alpha}\frac{\p}{\p\sigma^\beta}=\hbar^2\left(\frac{\p}{\p r}\right)^2+\left(\frac{\p}{\p\theta}\right)^2,
\ee
we see that $h$ in the tropical limit becomes also degenerate,
\be
h^{\alpha\beta}\frac{\p}{\p\sigma^\alpha}\frac{\p}{\p\sigma^\beta}=\left(\frac{\p}{\p\theta}\right)^2,
\ee
in such a way that $g$ and $h$ are no longer inverses of each other, but instead satisfy the ``mutual invisibility'' condition
\be
g_{\alpha\beta}h^{\beta\gamma}=0,\qquad h^{\alpha\beta}g_{\beta\gamma}=0.
\label{eemutinv}
\ee
We can set up an analog of the natural definition of conformal symmetry for $h^{\alpha\beta}$: Which diffeomorphisms of $\Sigma$ map this degenerate inverse metric to itself up to a $z=1$ Weyl transformation?  This condition in the covariant form is now given by
\be
-h^{\alpha\gamma}\p_\gamma\xi^\beta-h^{\beta\gamma}\p_\gamma\xi^\alpha=\tilde\omega(\sigma^\gamma)h^{\alpha\beta},
\ee
for some infinitesimal Weyl rescaling parameter $\tilde\omega$.  In components, these conditions reduce to
\be
\p_\theta\xi^r=0,\qquad -2\p_\theta\xi^\theta=\tilde\omega(r,\theta).
\ee
They are solved by
\be
\xi^r=f(r),\qquad \xi^\theta=-\frac{1}{2}\int \tilde\omega(r,\theta)\,d\theta,
\ee
with $\tilde\omega(r,\theta)$ arbitrary.  This symmetry algebra is again adapted to the foliation, and again larger than the symmetry algebra of the Jordan structure.  

\subsubsection{Weyl transformations and symmetries of the Jordan structure}

The symmetries of the Jordan structure will emerge as conformal symmetries of our degenerate metric structure when we allow for the presence of both $g_{\alpha\beta}$ and $h^{\alpha\beta}$, and require that they transform under the Weyl rescalings with the opposite weight.  This last condition requires $\tilde\omega (\sigma^\gamma)=-\omega(\sigma^\gamma)$, and one can show that under this requirement, the intersection of the conformal symmetry algebras of $g$ and $h$ treated separately results precisely in the symmetry algebra of the Jordan structure.  This gives a clear geometric interpretation of the symmetry algebra of the Jordan structure as a nonrelativistic conformal symmetry associated with isotropic Weyl rescalings of a pair $g_{\alpha\beta}$, $h^{\alpha\beta}$ satisfying the mutual invisibility condition mentioned above.

In retrospect, this result is in fact not geometrically very surprising.  Consider a Jordan structure $\varepsilon$ on $\Sigma$.  In the preferred local coordinates, $\varepsilon$ takes the form
\be
\varepsilon_\alpha{}^\beta d\sigma^\alpha\frac{\p}{\p\sigma^\beta}=dr\frac{\p}{\p\theta}.
\ee
Here $dr$ is an eigenvector (with eigenvalue zero) of $\varepsilon$ acting as an automorphism on $T^\ast\Sigma$, and $\p/\p\theta$ is similarly an eigenvector of $\varepsilon$ acting as an automorphism of $T\Sigma$.  This eigenvector $dr$ is determined by $\varepsilon$ uniquely, up to a Weyl transformation $dr\to \Omega(r,\theta)dr$, and $d\theta$ is similarly determined up to the Weyl transformation with the opposite weight, $\p/\p\theta\to \Omega^{-1}(r,\theta)\p/\p\theta$.  In turn, $dr$ and $\p/\p\theta$ define degenerate symmetric tensors
\be
g=dr^2,\qquad h=\left(\frac{\p}{\p\theta}\right)^2,
\ee
which are uniquely determined by $\varepsilon$ precisely up to the Weyl rescalings $g\to\Omega^2 g$ and $h\to\Omega^{-2}h$, respectively.  Thus, we see that the Jordan structure uniquely determines the degenerate pair $g$ and $h$ up to the Weyl transformations, and therefore the conformal symmetries of this pair coincide with the symmetries of the Jordan structure itself.

\subsection{Symmetries of the localization equations}
\label{sslocsym}

As we have seen in \S\ref{ssjsym}, the localization equations are invariant under our anisotropic conformal transformations, if $X$ and $\Theta$ transform as conformal scalars, as in (\ref{eeinfconfsc}).   Besides this conformal symmetry, the localization equations as written in a preferred conformal coordinate system $(r,\theta)$ 
\be
\p_rX-\p_\theta\Theta=0,\qquad \p_\theta X=0,
\ee
exhibit an additional important symmetry, independent for each $(X,\Theta)$ pair.  The first equation is clearly invariant under
\bea
\delta X(r,\theta)&=&\p_\theta\alpha(r,\theta),\\
\delta \Theta(r,\theta)&=&\p_r\alpha(r,\theta),
\eea
for arbitrary $\alpha(r,\theta)$.  Requiring that the second equation also be invariant restricts $\alpha$ to be at most linear in $\theta$, and we obtain an intriguing symmetry
\bea
\delta X(r,\theta)&=&\alpha_1(r),\\
\delta \Theta(r,\theta)&=&\alpha_0(r)+\theta\p_r\alpha_1(r),
\eea
with $\alpha_0(r)$ and $\alpha_1(r)$ arbitrary projectable functions on the foliation,
\ie , functions of only $r$.

At first glance, the precise status of such a symmetry appears a bit mysterious.  It looks like a hybrid between a linear shift symmetry in $\theta$, while it still exhibits an arbitrary dependence on $r$.  If we interpret $r$ as a time variable, this would perhaps suggest an underlying gauge invariance, and associated constraints on the momenta in the canonical quantization.  

\subsection{Admissible singularities in the foliation}
\label{ssadmis}

When we attempt to define the path integral for the tropological sigma model on a higher-genus worldsheet $\Sigma$, inevitably some leaves of the foliation induced by the Jordan structure must be singular.  This signals the presence of controllable topological defects in the gravity sector of the worldsheet theory.  In this paper, we consider the Jordan structure as a non-dynamical given.  This still allows us to consider several natural types of defects in the Jordan structure and investigate whether the localization equations continue making sense in the vicinity of such singularities of the foliation.  For this, it is important that we have our proposed localization equations written in the covariant form, which does not require the existence of the adapted coordinates $(r,\theta)$ around the singular points, where no adapted coordinate systems exist.

Some of the simplest examples of the singular leaves in the foliation, and the associated topological defects in the Jordan structure, are depicted in Figure~\ref{ffdefect}. 

\begin{figure}[t!]
  \centering
    \includegraphics[width=0.7\textwidth]{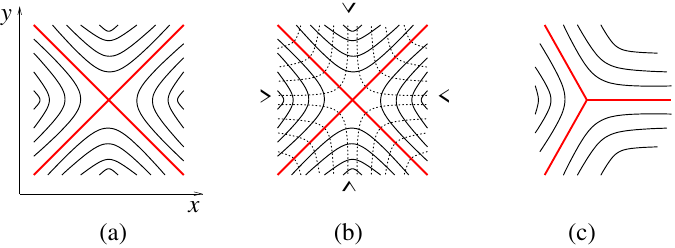}
    \caption{Defects in the Jordan structure lead to singular leaves in the associated foliation.}
    \label{ffdefect}
\end{figure}

In physics terms, the singularities in our foliations of the worldsheet should be viewed as topological defects of the dynamical Jordan structure.  While the full analysis of the allowed defects and their dynamics would require a more detailed description of the dynamics in the worldsheet gravity sector (which will be developed in the sequel \cite{sequel}), in this paper we can at least consider the topological classification of such defects, and check to what extent are the localization equations for the sigma model still satisfied near such defects.

First, note that the natural order parameter that classifies singular leaves in the foliation is essentially the same as the order parameter in two-dimensional nematic liquid crystals:  It is represented by an unoriented direction in the tangent space to the two-dimensional surface.  In our case, it is natural to orient this order parameter along the leaves of the foliation.  In the nematic liquid crystal case, the order-parameter manifold is an $S^1$, implying that the corresponding homotopy groups classifying stable defects are such that the point-like defects on the plane are labeled by one integer, the winding number of the unoriented direction as we circumnavigate around the point-like defect.  In our worldsheets with foliations, these types of defects have been also encountered frequently in the mathematical literature on foliated manifolds.  Consider a candidate Jordan structure on a compact surface $\Sigma$, of genus $g>1$.  The induced foliation will inevitably have some singular leaves.  We will choose the Jordan structure such that the number of such singular leaves is finite, and such that the rest of $\Sigma$ consists of a finite number of sleeves ending with their boundary components on the singular leaves.

\subsubsection{Junctions of sleeves}

A typical such configuration is locally depicted in Figure~\ref{ffdefect}(a), with four sleeves meeting at the singular leaf.  We will focus for simplicity on discussing this simple case, but the construction is analogous for any finite number of sleeves meeting.  In Figure~\ref{ffdefect}(a), nonsingular leaves inside each sleeve are indicated by the curved lines.  We wish to investigate suitable junction conditions that such sleeves must satisfy at the singular leaf; this is going to be important for our ability to count the number of instantons contributing to the path integral of the tropological sigma models on higher-genus worldsheets $\Sigma$.  We will describe the local geometry near the singular point of the singular leaf using an atlas containing five coordinate charts.  We will introduce coordinates $x,y$ such that the singular leaf is locally described by $x=\pm y$, and in particular the coordinate system is well-defined at the singular point of the foliation at $x=y=0$.  These are \textit{not} adapted coordinates to the foliation, but are a perfectly legitimate coordinate choice on $\Sigma$ interpreted as a differentiable manifold, with $x=y=0$ a smooth point.  Next, on the internal portion of each of the four sleeves we introduce natural adapted coordinates.  We label the four sleeves by $>,<,\wedge,\vee$ correspondingly (see Figure~\ref{ffdefect}(b)).  On the first sleeve, we choose adapted coordinates $r_<,\theta_<$, such that $r_<$ is negative inside the sleeve, and approaches zero as we approach the singular leaf.  Similarly, $\theta_<$ will represent a coordinate along the leaves of the sleeve, chosen such that $\theta_<=0$ along the $y=0$ axis.  Analogous adapted coordinate systems are chosen on the remaining three sleeves.  Each of these coordinate systems can be naturally continued inside the neighboring two sleeves, leading to the natural transition functions between them.  For instance, $r_<$ can be continued to positive values, which for $\theta_<$ positive extends smoothly into the neighboring $\vee$ sleeve (and for $\theta_<$ negative into the $\wedge$ sleeve), with natural transition functions 
\bea
r_\vee=-r_<,&&\qquad \theta_\vee=-\theta_<,\cr
r_\wedge=-r_<,&&\qquad \theta_\wedge=-\theta_<,
\eea
and similarly for all the other neighboring pairs of sleeves.

These four adapted charts cover the open domain of $\Sigma$ around the $x=y=0$ singular point, with this point removed.  In order to complete the description around the singular point, we must specify the transition functions from the four adapted coordinate systems inside the sleeves to the $x,y$ coordinates.  On the overlap with the $r_<,\theta_<$ coordinate system, one can for instance take
\be
r_<=y^2-x^2,\qquad \theta_<=2xy,
\ee
and similarly for the remaining overlaps.  (The lines of constant $\theta_<$ are indicated in Figure~\ref{ffdefect}(b) as dotted lines.)

This establishes an atlas of smooth charts in the vicinity of the defect in the Jordan structure.

A few comments are in order:

(i) Note that the open portions of the singular leaf of the foliation away from $x=y=0$ look locally smooth, and indistinguishable from local neighborhoods in regular $S^1$ leaves of the foliation.  The entire singular feature of the singular leaf is associated with the juncture point at $x=y=0$.

(ii) Having found a suitable coordinate description of the vicinity of the singular point, we can now evaluate the Jordan structure in the coordinates $x,y$ suitable for taking the $x,y\to 0$ limit.  We get:
\be
\varepsilon_\alpha{}^\beta=\frac{1}{x^2+y^2}\begin{pmatrix} xy & x^2\\ -y^2&-xy\end{pmatrix}.
\label{eefoureps}
\ee
We see that this Jordan structure is indeed singular at the origin in the $x,y$ coordinate system:  While the values of the components $\varepsilon_\alpha{}^\beta$ obtained as $x$ and $y$ are taken simultaneously to zero are finite, the result depends on the angle with which the singular point is approached.

(iii) Fortunately, for the calculations in the path integral of the tropological sigma model, we do not need the Jordan structure to be nonsingular at the singular point of its foliation, it is sufficient if the localization equations are unambiguously defined, and satisfied, near such singular points.  The localization equations in the $x,y$ coordinate system take the following form,
\bea
\frac{1}{x^2+y^2}\begin{pmatrix} xy & x^2\\ -y^2&-xy\end{pmatrix}\begin{pmatrix}\p_x \Theta\\ \p_y \Theta\end{pmatrix}+\begin{pmatrix}\p_x X\\ \p_y X\end{pmatrix}&=&0,\\
\frac{1}{x^2+y^2}\begin{pmatrix} xy & x^2\\ -y^2&-xy\end{pmatrix}\begin{pmatrix}\p_x X\\ \p_y X\end{pmatrix}&=&0,
\eea
and they are naturally solved by maps that have a second-order zero at $x=y=0$, 
\be
X=C(y^2-x^2),\qquad \Theta=2Cxy,
\ee
for some constant $C$.  This compensates for the singularity of the Jordan structure, and makes the continuation of the solutions of the localization equations unambiguous at the juncture point in the singular leaf.  Thus, it appears natural that such mild singularities in the worldsheet Jordan structure should be admissible; they are indeed inevitable for constructing solutions of the localization equations on higher-genus surfaces.

The construction is easily generalized to the case with other values of the defect quantum number around a juncture in the singular leaf.  An example with three sleeves meeting at a singular leaf is depicted in Figure~\ref{ffdefect}(c).  The main novelty in the case when an odd number of sleeves meet at the singular point of the singular leaf is the fact that $X$ and $\Theta$ are then antiperiodic as one circumnavigates the singular foliation point.  In general, any number of sleeves greater than two can meet at a singular leaf.

Having understood the local geometric structure of worldsheets near singular points of the foliation, we can now construct global solutions of the localization equations, by gluing together the sleeve solutions that we found in (\ref{eeglobx}), (\ref{eeglobth}) (or their more general, local form (\ref{eelocx}), \ref{eelocth})) at the singular leaves where several sleeves meet.  This is accomplished with the use of the $\alpha$ symmetries discovered in \S\ref{sslocsym}.  Take for example the global solutions on two neighboring sleeves,
\bea
X_<=n_<r_<,&&\qquad \Theta_<=\Theta_{<0}(r)+n_<\theta_<,\\
X_\vee =n_\vee r_\vee, &&\qquad \Theta_\vee =\Theta_{\vee 0}(r)+n_\vee\theta_\vee.
\eea
We will consider the nondegenerate case, when $n_<$ and $n_\vee$ are both nonzero integers.
To match such solutions on the coordinate patch where they overlap, one can use the $\alpha_0(r)$ symmetry to set $\Theta_{<0}=\Theta_{\vee 0}$ on the overlap, and then use the $\alpha_1(r)$ symmetry to rescale $r$ such that $n_<=n_\vee$ in the local vicinity of the singular point.
(If one of the integers $n_<,$ $\ldots,$ is zero, the corresponding sleeve is entirely mapped to a marked point modulo an $\alpha$ transformation, and can therefore be collapsed to a worldsheet point.)

Iterating this process for all pairs of neighboring sleeves, one constructs a global solution of the localization equation, assuming that a single global topological constraint is satisfied for each singular leaf of the foliation:  The map from $\Sigma$ to $M$ will be continuous (and, indeed, smooth) if at each singular leaf the total winding numbers of all the attached sleeves sum up to zero.  Thus, for example, when four leaves meet at one singular leaf (as locally depicted in Figure~\ref{ffdefect}(a)), they must globally satisfy
\be
n_<+n_\wedge+n_>+n_\vee=0.
\ee
If the target space has more than one $X^I,\Theta^I$ pairs of coordinates, there is one such condition for each pair, at each singular leaf.  This precisely reproduces our topological expectations, and matches for example the analogous charge conservation conditions at Type IIB string junctions.  

\subsubsection{Punctures}
\label{sspunct}

Finally, one last type of singularity in the worldsheet Jordan structure should be discussed:  a puncture (see Figure~\ref{ffpuncture}).  We use it to simply indicate the presence of a semi-infinite sleeve, whose natural adapted $r$ coordinate goes to $\infty$ (or $-\infty$).  We find it useful to depict it simply as a marked point on $\Sigma$ (which one can think of as representing a singular leaf of the worldsheet foliation, ``at infinity''), with the understanding that the coordinates that would be smooth around such a marked point do not belong to the class of adapted coordinates of the Jordan structure.  For example, one can relate the adapted coordinates $(r,\theta)$ to coordinates $(x,y)$, well-defined at the puncture at $r\to-\infty$, via
\be
x=e^r\cos\theta,\qquad y=e^r\sin\theta.
\ee
Then one finds that the standard Jordan structure on the semi-infinite sleeve is given in the $(x,y)$ coordinates by
\be
\varepsilon_\alpha{}^\beta=\frac{1}{x^2+y^2}\begin{pmatrix} -xy & x^2\\ -y^2&xy\end{pmatrix}.
\ee
In these non-adapted coordinates at the puncture, these components of the Jordan structure look superficially similar as those at the singular leaf where four sleeves meet, (\ref{eefoureps}).  However, despite this superficial similarity, the solutions of the localization equations in the $(x,y)$ coordinates around the puncture do not exhibit zeros, but instead they diverge at the puncture, 
\bea
X(x,y)&=&\frac{n}{2}\log (x^2+y^2),\\
\Theta(x,y)&=&n \arctan\left(\frac{y}{x}\right),
\eea
indicating that the end of the sleeve (when $n\neq 0$) is indeed at infinity.

\begin{figure}[t!]
  \centering
    \includegraphics[width=0.4\textwidth]{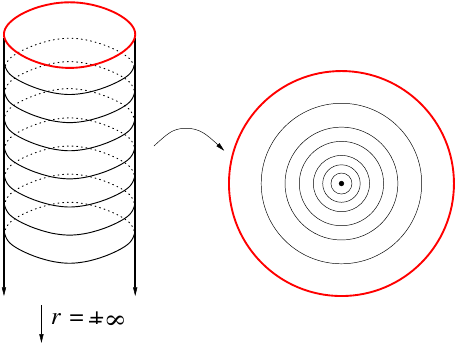}
    \caption{It is often convenient to depict the half-infinite sleeve as a punctured disk, even though the local coordinates around the marked point on the disk are not adapted to the Jordan structure.}
    \label{ffpuncture}
\end{figure}

Does this singular behavior of $X$ and $\Theta$ at the puncture mean that punctures should be treated as singularities not belonging to the worldsheet $\Sigma$?  In fact, the opposite turns out to be true: The punctures can be consistently treated as smooth points in the compact worldsheet $\Sigma$, mapped smoothly to the singular foliation leaves of the target space.  In order to see that, one needs to switch not only to the $(x,y)$ coordinates near the worldsheet puncture, but also to similar non-adapted coordinates $(\CX,\CY)$ that cover a neighborhood of the singular leaf at $X=-\infty$ in the target $\MT P^1$.  We define
\bea
\CX&=&e^X\cos\Theta,\\
\CY&=&e^X\sin\Theta.
\eea
In such non-adapted coordinates defined in the vicinity of the singular leaf at $X=0$, the standard Jordan structure on the $\MT P^1$ takes the form
\be
J_i{}^j=\frac{1}{\CX^2+\CY^2}\begin{pmatrix} -\CX\CY & \CX^2\\ -\CY^2&\CX\CY\end{pmatrix}.
\ee
When expressed in this pair of non-adapted coordinate systems $(x,y)$ on the worldsheet and $(\CX,\CY)$ on the target space, the localization equations
\be
\frac{1}{x^2+y^2}\begin{pmatrix} -xy & x^2\\ -y^2&xy\end{pmatrix}
  \begin{pmatrix}\p_x\CX&\p_x\CY\\
\p_y\CX&\p_y\CY\end{pmatrix}
-\frac{1}{\CX^2+\CY^2}\begin{pmatrix}\p_x\CX&\p_x\CY\\
  \p_y\CX&\p_y\CY\end{pmatrix}\begin{pmatrix} -\CX\CY & \CX^2\\ -\CY^2&\CX\CY\end{pmatrix}=0
    \label{eelocxy}
\ee
look quite complicated, but they and their solutions can be successfully continued through the puncture at $x=y=0$.  In fact, there is an infinite sequence of natural solutions to these nonlinear equations, given by simple monomials in $x$ and $y$ of arbitrarily high degree $n$,
\bea
n=1:&&\qquad\CX=x,\qquad\qquad\ \ \CY=y,\nonumber\\
n=2:&&\qquad\CX=x^2-y^2,\quad\quad\CY=2xy,\\
n=3:&&\qquad\CX=x^3-3xy^2,\ \ \ \CY=-y^3+3x^2y,\nonumber\\
&\vdots&\qquad.\nonumber
\eea
At higher integer values of $n$, the structure of these monomial solutions can best be illustrated as follows.  Combine the worldsheet and target-space coordinates into complex coordinates,
\be
z=x+iy,\qquad \CZ=\CX+i\CY.
\ee
(This is not meant to imply the existence of any standard complex structure near the worldsheet puncture or near the singular leaf of the $\MT P^1$ target space -- these complex coordinates are introduced merely for convenience.)  In these complex coordinates, the localization equations (\ref{eelocxy}) boil down to
\bea
2\p_{\bar z}\CZ-\frac{z}{\bar z}\p_z \CZ+\frac{\CZ}{\bar\CZ}\p_{\bar z}\bar\CZ&=&0,\\
\frac{\bar z}{z}\p_{\bar z}\CZ+\frac{\CZ}{\bar\CZ}\p_z\bar\CZ&=&0.
\eea
Clearly, these nonlinear equations are not the equations for holomorphicity in standard complex geometry; yet, remarkably, they have an infinite sequence of solutions given simply by holomorphic monomials,
\be
\CZ_{(n)}(z,\bar z)=z^n.
\ee
These monomial solutions labeled by $n$ represent smooth continuations of the standard sleeve solutions (\ref{eeglobx}), (\ref{eeglobth}) with winding number $n$ through the puncture.  As we see from this analytic form, the localization equations are again smoothly extended through the singular leaf of the foliation at the puncture, and we have a good notion of smoothness at the puncture for the maps to the target space.

\subsection{Tropical limit as a nonrelativistic limit}

In the context of the Litvinov-Maslov dequantization, the idea of the tropical limit emerged in attempts to make sense of the classical asymptotics of quantum systems and their observables.  In our two-dimensional topological sigma model context, we have found that the ``dequantization'' of the tropical geometry does not correspond to the classical limit of our system; instead, the $\hbar\to 0$ limit of the Litvinov-Maslov dequantization corresponds to a \textit{nonrelativistic limit}, in which the worldsheet theory undergoes a $c\to\infty$ contraction, with $c$ the worldsheet speed of light.  In this limit, the leaves of the foliation represent the worldsheet spatial dimension, and the direction transverse to the leaves is the worldsheet time.  As a result, the symmetries of the tropical geometry naturally acquire a field-theoretical interpretation as nonrelativistic symmetries of the worldsheet theory.

Our quantum topological field theory of the tropological sigma model will of course have its own Planck constant, which will be denoted by $e$ below.  This Planck constant has nothing to do with the $\hbar$ of the Litvinov-Maslov dequantization.  Based on the standard BRST arguments, the semiclassical approximation $e\to 0$ will be exact in the path integral of our tropological sigma models.  

\section{Tropological sigma models}

Having understood the basic geometric features of Jordan structures, and its symmetries, we are now equipped to construct a path integral formulation of our theory using the methods of topological field theories of the cohomological type \cite{ewcoho}.  In this theory, the worldsheet path integral for the correlation functions of physical observables will localize to the solutions of our localization equations representing tropicalization.

\subsection{Cohomological BRST construction from the localization equations}

We follow the standard logic of cohomological field theories, and use the traditional methods of BRST quantization to construct the path integral \cite{ewcoho}.  All fields of the BRST quantization fall into multiplets of the BRST charge $Q$, which satisfies $Q^2=0$ and defines physical observables through its cohomology.    

\subsubsection{Ghosts}

Our basic BRST multiplet reflects the underlying bosonic topological symmetry (\ref{eetopofs}), transforming $Y^i$ into the corresponding ghost $\psi^i$: 

\be
[Q,Y^i]=\psi^i.
\ee
When we use the adapted coordinates $X$ and $\Theta$ on the target space, we will use the following simplified notation for the individual components of the ghost field, 
\be
\psi^X\equiv\psi,\qquad \psi^\Theta\equiv\Psi.
\ee
Thus, in the adapted coordinates the basic BRST multiplet associated with each tropicalized complex dimension of the target space is given by
\bea
[Q,X]&=&\psi,\qquad \{Q,\psi\}=0,\\
{[Q,\Theta]}&=&\Psi,\qquad\{Q,\Psi\}=0.
\eea

\subsubsection{Antighosts and auxiliaries}
\label{ssanti}

In order to perform the gauge-fixing of the topological symmetry using the localization equation $E_\alpha{}^i=0$, we introduce the trivial BRST multiplet
\bea
\{ Q,\chi^\alpha{}_i\}&=&\CB^\alpha{}_i,\\
{[Q,\CB^\alpha{}_i]}&=&0,
\eea
containing the bosonic auxiliary field $\CB^\alpha{}_i$ and its antighost superpartner $\chi^\alpha{}_i$ which must be chosen such that the integral
\be
\int_\Sigma d^2\sigma\,\CB^\alpha{}_i\,E_\alpha{}^i
\label{eebelag}
\ee
is covariantly well-defined.  In addition, it is often customary to introduce another BRST exact term in the action, which is quadratic and nondegenerate in the auxiliaries $\CB^\alpha{}_i$, so that they can be integrated out by Gaussian integration.  This in turn leads to the bosonic part of the action that is nondegenerate and quadratic in $E_\alpha{}^i$.  

When we try to repeat this strategy in the tropical case, we encounter several intriguing subtleties, which we now address.  Note first that since $E_\alpha{}^i$ is a section of $T^\ast\Sigma\otimes\Phi^\ast(TM)$, $\CB^\alpha{}_i$ must be a density-valued section of the dual bundle $T\Sigma\otimes\Phi^\ast(T^\ast M)$.  In addition, since $E_\alpha{}^i$ satisfies (\ref{eetrself}) and therefore has only two independent components, only two out of the four components of $\CB^\alpha{}_i$ will couple.  How can we formulate this reduction of $\CB$ in geometrical terms, and can it be done covariantly?

In the relativistic case, one uses the fact that every tensor $\CB^\alpha{}_i$ can be decomposed into its self-dual and anti-self dual part, and simply restricts $\CB^\alpha{}_i$ to be (anti)-self-dual.  This reduces the number of independent components to two, which are the appropriate ones that couple to the two components of the localization equation.  In our case, one can again define the natural tropicalized limit of (anti)-self-duality equations,
\be
\varepsilon_\beta{}^\alpha\CB_\pm^\beta{}_i=\pm J_i{}^k\CB_\pm^\alpha{}_k.
\label{eebselfd}
\ee
However, declaring the auxiliary field to be self-dual or anti-self-dual in this way does not solve our problem:  A quick inspection in adapted coordinates shows that either of the two choices throws away one of the components of $\CB^\alpha{}_i$ that needs to couple to $E_\alpha{}^i$.  This is again related to the observations about vector spaces with Jordan structures that we made above:  The structures induced on the vector spaces are often filtrations, not direct-sum decompositions.  As a result, tensors of the type $\CB^\alpha{}_i$ do not decompose into a sum of $\CB_+^\alpha{}i$ and $\CB_-^\alpha{}_i$ that satisfy (\ref{eebselfd}) for the two corresponding sign choices.  The reasons can be seen in the adapted coordinates -- first of all, both equations in (\ref{eebselfd}) require $\CB^r{}_\Theta=0$, so only those tensors that have this component vanishing can be written as a sum of solutions of (\ref{eebselfd}).  But even then, such a decomposition will not be unique: The $\CB_+^\theta{}_\Theta$ and $\CB_-^\theta{}_\Theta$ components are completely unrestricted by (\ref{eebselfd}), and setting their sum equal to $\CB^\theta{}_\Theta$ leaves an ambiguity.  In addition, $\CB^r{}_\Theta$ is precisely one of the components that we need to keep around, since $E_r{}^\Theta$ is one of the two nonzero components in the localization equation.

A fully covariant way how to achieve the desired reduction of $\CB^\alpha{}_i$ does exist, but it requires the introduction of an additional gauge symmetry.  Observe that our action (\ref{eebelag}) exhibits a gauge invariance under 
\be
\delta\CB^\alpha{}_i=f_+^\alpha{}_i(\sigma^\beta),\qquad \delta Y^i=0,
\label{eegaugeplus}
\ee
where $f_+^\alpha{}_i(\sigma^\beta)$ satisfies the tropicalized self-duality condition (\ref{eebselfd}) with the plus sign choice.  Thus, the covariant reduction of the four components in $\CB^\alpha{}_i$ cannot be achieved as a reduction to a two-dimensional subspace, but instead, it is covariantly represented as a two-dimensional coset space, defined modulo the gauge transformations (\ref{eegaugeplus}).

The next question to ask is whether -- at least in the adapted coordinates -- there is a natural gauge-fixing choice for our new gauge symmetry (\ref{eegaugeplus}).  One simple choice would seem to suggest itself:
\be
\CB^\theta{}_X=0,\qquad \CB^r{}_X=-\CB^\theta{}_\Theta.
\label{eenaiveg}
\ee
This is certainly a fine gauge-fixing choice for (\ref{eegaugeplus}), and it appears to mimic the symmetry structure of $E_\alpha{}^i$ that the auxiliaries are coupling to.  However, perhaps somewhat surprisingly, this gauge choice turns out to violate the worldsheet conformal invariance!

In order to find another gauge choice that is consistent with our worldsheet conformal invariance, we must abandon the full covariance in the target space.  Recall that tropicalized target spaces have no continuous symmetries that would mix various coordinates; in particular, there is no continuous symmetry that would transform the preferred coordinates $X$ and $\Theta$ into each other.  Until now, we often used more general coordinates $Y^i$, and implicitly required invariance of our formulas under arbitrary coordinate changes of $Y^i$.  Once we are willing to use the adapted coordinates $X$ and $\Theta$ on the target space, it turns out that there is a unique natural gauge-fixing condition, which is still fully covariant under arbitrary coordinate changes on $\Sigma$,
\be
\CB^\alpha{}_X=0.
\ee
As we will see below, this condition is also uniquely determined by requiring the consistency with worldsheet conformal transformations.  From now on, we adopt this gauge-fixing condition, and formulate the theory using the adapted coordinates $X,\Theta$ on the target.  When we also use the adapted coordinates $r,\theta$ on $\Sigma$, we will simply refer to the two remaining components of $\CB^\alpha{}_i$ using the following simpler notation:
\be
\CB^r{}_\Theta\equiv B,\qquad\CB^\theta{}_\Theta\equiv -\beta.
\ee

One additional subtlety appears when we attempt to add a term to the action quadratic in $B$ and $\beta$, to integrate them out.  In the worldsheet covariant form, such a term would be given by
\be
-\frac{1}{2}\int d^2\sigma\,\gamma_{\alpha\beta}H^{\Theta\Theta}\CB^\alpha{}_\Theta\CB^\beta{}_\Theta.
\ee
Here $H^{\Theta\Theta}$ is the component of the degenerate inverse metric on the target in the tropical limit, which we have indeed found to be non-zero.  Finally, Since each $\CB^\alpha{}_\Theta$ transforms under worldsheet diffeomorphisms as a tensor density, $\gamma_{\alpha\beta}$ must itself be a tensor density of weight $-1$.

In the relativistic case, there would be a natural candidate
\be
\hat\gamma_{\alpha\beta}=\frac{1}{\sqrt{\hat g}}\hat g_{\alpha\beta},
\label{eegammat}
\ee
where $\hat g_{\alpha\beta}$ is a nondegenerate worldsheet metric, and $\hat g$ its determinant.  This relativistic $\hat\gamma_{\alpha\beta}$ is sensitive only to the relativistic conformal structure defined by $\hat g_{\alpha\beta}$.  In our tropical case, we do not have a nondegenerate worldsheet metric, only its degenerate limit $g_{\alpha\beta}$.  Similarly, there is a degenerate limit $\gamma_{\alpha\beta}$ of (\ref{eegammat}), whose only non-zero component is $\gamma_{rr}=1$, much as the only nonzero component of $g_{\alpha\beta}$ in the adapted coordinates is $g_{rr}$.  This $\gamma_{\alpha\beta}$ can be used to covariantly square one component of $\CB^\alpha{}_\Theta$, but not the other.   Thus, we can add a BRST invariant and conformally invariant term to our action, given in adapted coordinates simply by
\be
S=-\frac{1}{2}\int dt\,d\theta\,B^2.
\ee
$B$ can then be integrated out by the Gaussian integral, to simplify the action; however, $\beta$ cannot acquire a conformally invariant quadratic term, and therefore the action will stay linearly dependent on $\beta$ if we wish to maintain our nonrelativistic conformal invariance of the full BRST quantized theory.

The structure of the antighosts $\chi^\alpha{}_i$ now follows the same pattern as that of the auxiliaries.  Only two components of this fermionic tensor will couple to the ghosts $\psi^i$.  In the fully covariant formulation, there is a fermionic superpartner symmetry acting on the antighosts,
\be
\delta\chi^\alpha{}_i=\varphi_+^\alpha{}_i(\sigma^\beta),
\ee
where $\varphi_+^\alpha{}_i$ is a fermionic analog of $f_+^\alpha{}_i$ satisfying the same equation (\ref{eebselfd}).  It can be uniquely gauge-fixed in a way consistent with worldsheet conformal invariance by setting
\be
\chi^\alpha{}_X=0,
\ee
again using the adapted coordinates $X$ and $\Theta$ on the target.

If we in addition choose to use the adapted coordinates $r,\theta$ on the worldsheet, the two remaining components of the antighost will be simply denoted by
\be
\chi^r{}_\Theta\equiv\Chi,\qquad \chi^\theta{}_\Theta\equiv -\chi.
\ee
(We suggest pronouncing the symbol for the antighost $\Chi$ as ``capital chi''.)

Having sorted out the structure of our BRST multiplets, we are now ready to construct the path integral using the standard methods of BRST quantization.

\subsubsection{Lagrangians}

We will define the tropological sigma model in terms of a path integral,
\be
\int\CD X\,\CD \Theta\,\CD\psi\,\CD\Psi\,\CD\chi\,\CD\Chi\,\CD B\,\CD\beta\, e^{-S},
\ee
with the action $S$ being a worldsheet integral of a local Lagrangian which is BRST exact, modulo possibly adding topological terms:
\bea
S&=&\frac{1}{e}\int dr\,d\theta\left\{ Q,\Chi(\p_\theta\Theta-\p_r X -\textstyle{\frac{1}{2}}B)+\chi\,\p_\theta X\right\}\nonumber\\
&&{}=\frac{1}{e}\int dr\,d\theta\left\{B(\p_\theta\Theta-\p_r X)+\beta\p_\theta X-\frac{1}{2} B^2
-\Chi(\p_\theta\Psi-\p_r\psi)-\chi\p_\theta\psi\right\}.
\eea
As promised, we have introduced the Planck constant of our quantum theory, $e$.  By the standard BRST arguments of cohomological field theories, the semiclassical one-loop approximation in $e$ will be exact, leading to the localization of the path integral to the solutions of the localization equations.

We can integrate out $B$, to bring the action to a more standard form,
\be
S=\frac{1}{e}\int dr\,d\theta\left\{\frac{1}{2}\left(\p_\theta\Theta-\p_r X\right)^2+\beta\p_\theta X-\Chi(\p_\theta\Psi-\p_r\psi)-\chi\p_\theta\psi\right\}.
\label{eeactquad}
\ee
This action is invariant under the BRST transformations
\bea
[Q,X]&=&\psi,\qquad \{Q,\psi\}=0,\\
{[Q,\Theta]}&=&\Psi,\qquad\{Q,\Psi\}=0,\\
\{Q,\Chi\}&=&\p_\theta\Theta-\p_r X,\\
\{Q,\chi\}&=&\beta,\qquad [Q,\beta]=0.
\eea
Both actions are also invariant under the nonrelativistic conformal transformations.  Before $B$ has been integrated out, the conformal symmetries act via
\bea
\delta X&=&f(r)\p_r X+\left(\vphantom{\frac{1}{2}}F(r)+\theta\p_r f(r)\right)\p_\theta X,\\
\delta \Theta&=&f(r)\p_r\Theta+\left(\vphantom{\frac{1}{2}}F(r)+\theta\p_r f(r)\right)\p_\theta \Theta,\\
\delta B&=&f(r)\p_r B+B\p_rf(r)+\left(\vphantom{\frac{1}{2}}F(r)+\theta\p_r f(r)\right)\p_\theta B,\\
\delta\beta&=&f(r)\p_r\beta+\beta\p_rf(r)+\left(\vphantom{\frac{1}{2}}F(r)+\theta\p_r f(r)\right)\p_\theta\beta+B\,\p_r\left(\vphantom{\frac{1}{2}}F(r)+\theta\p_r f(r)\right).
\eea
Note that $B$ and $\beta$ transform in a way reminiscent of a Jordan pair, with $B$ transforming through terms involving only $B$, while $\beta$ transforms via terms involving both $\beta$ and $B$.  When $B$ is integrated out, the transformation rule for $\beta$ becomes dependent on $X$ and $\Theta$,
\be
\delta\beta=f(r)\p_r\beta+\beta\p_rf(r)+\left(\vphantom{\frac{1}{2}}F(r)+\theta\p_r f(r)\right)\p_\theta\beta+(\p_\theta\Theta-\p_r X)\p_r\left(\vphantom{\frac{1}{2}}F(r)+\theta\p_r f(r)\right).
\ee
The action (\ref{eeactquad}) is then consistently invariant under these reduced conformal transformations not involving $B$.  

For completeness, we list the conformal transformation properties of the remaining two components of $\CB^\alpha{}_i$ that we dropped by our gauge choice $\CB^\alpha{}_X=0$ for the gauge symmetry generated by $f_+^\alpha{}_i$ in \S\ref{ssanti}:
\bea
\delta\CB^r{}_X&=&f(r)\p_r \CB^r{}_X+\CB^r{}_X\p_rf(r)+\left(\vphantom{\frac{1}{2}}F(r)+\theta\p_r f(r)\right)\p_\theta \CB^r{}_X,\nonumber\\
\delta\CB^\theta{}_X&=&f(r)\p_r\CB^\theta{}_X+\CB^\theta{}_X\p_rf(r)+\left(\vphantom{\frac{1}{2}}F(r)+\theta\p_r f(r)\right)\p_\theta\CB^\theta{}_X-\CB^r{}_X\p_r\left(\vphantom{\frac{1}{2}}F(r)+\theta\p_r f(r)\right).\nonumber
\eea
We see that they transform into each other, again as a Jordan pair: $\CB^r{}_X$ into itself, and $\CB^\theta{}_X$ into itself and $\CB^r{}_X$.  By inspection of these transformation rules, one can now explicitly confirm several points we made above: (i) It is consistent with worldsheet conformal invariance to fix the $f_+$ gauge symmetry by setting $\CB^\alpha{}_X=0$; (ii) our naive first choice for the gauge-fixing condition, (\ref{eenaiveg}), would be inconsistent with conformal invariance, and (iii) while the $B^2$ term in the action is conformally invariant, no linear combination of $\beta^2$ and $\beta B$ is consistent with the conformal symmetries.

Compared to the standard relativistic case, the action of our tropological sigma model exhibits one additional novelty that needs to be addressed.  Unlike its relativistic counterpart, our action has residual gauge symmetries, closely related to the residual symmetries of the localization equations that we found in \S\ref{sslocsym}.  Consider a local simply-connected neighborhood $\CU$, with adapted coordinates $r,\theta$.  The bosonic sector of (\ref{eeactquad}) is invariant under
\bea
\delta X&=&\p_\theta\alpha,\\
\delta \Theta&=&\p_r\alpha,\qquad\delta \beta=0,
\eea
if the gauge parameter $\alpha(r,\theta)$ is further constrained to satisfy
\be
\p_\theta{}^2\alpha=0.
\ee
Similarly, the fermionic sector is invariant under a fermionic gauge transformation with Grassmannian gauge parameter $\zeta(r,\theta)$,
\bea
\delta\psi&=&\p_\theta\zeta,\qquad \delta\Chi=0,\\
\delta\Psi&=&\p_r\zeta,\qquad \delta\chi=0,
\eea
if $\zeta(r,\theta)$ is constrained by
\be
\p_\theta{}^2\zeta=0.
\ee
The constraint equations on the gauge parameters are solved by $r$-dependent affine functions of $\theta$,
\be
\alpha=\alpha_0(r)+\theta\alpha_1(r),\quad\zeta=\zeta_0(r)+\theta\zeta_1(r),
\ee
yielding the gauge symmetries in the unconstrained form
\bea
\delta X&=&\alpha_1(r),\nonumber\\
\delta \Theta&=&\p_r\alpha_0(r)+\theta\p_r\alpha_1(r),\nonumber\\
\delta\psi&=&\zeta_1(r),\\
\delta\Psi&=&\p_r\zeta_0(r)+\theta\p_r\zeta_1(r),\nonumber\\
\delta \beta&=&\delta\Chi=\delta\chi=0.\nonumber
\eea
First, a few comments about the nature of these symmetries:

(i) The $\alpha$ and $\zeta$ symmetries are a hybrid between a genuine gauge redundancy with an arbitrary dependence on worldsheet coordinates (here exhibited only along $r$), and a global linear shift symmetry (along $\theta$):  The independent parameters $\alpha_0$, $\alpha_1$ and $\zeta_0$, $\zeta_1$ are projectable functions on the worldsheet foliation, a phenomenon that has no analog in relativistic theories.

(ii) If we interpret $r$ as (imaginary) worldsheet time -- and that is the interpretation here, in the topological context -- then $\alpha$ indeed represents a true gauge redundancy, with constraints on the canonical variables in the canonical quantization along $r$, and a non-uniqueness of the time evolution of a classical solution given fixed initial conditions.  If we studied the theory in the cross-channel, using $\theta$ as time, the interpretation of the $\alpha$ symmetries would be less clear.

(iii) If $\alpha$ were not constrained to be linear in $\theta$, its action on $X$ and $\Theta$ would be as the gauge symmetry in relativistic electromagnetism in two dimensions, with $X$ and $\Theta$ playing the role of the components of the electromagnetic gauge field.  However, this analogy is incomplete for two reasons: First, our action also contains the $\beta$-dependent term (which in the electromagnetic analogy looks a little like a gauge-fixing term), and perhaps more importantly, our $\Theta$ is a periodic variable with periodicity $2\pi$.

How should these residual gauge symmetries be treated, and do they require a secondary BRST gauge fixing, leading to a second generation of ``ghosts-for-ghosts''?  If we decided to treat these symmetries as secondary, in addition to the original topological symmetries (\ref{eetopofs}), it would be natural to implement them equivariantly, as for example is the case with the ordinary Yang-Mills gauge symmetries in topological Yang-Mills theories.  This would in turn lead to second-generation bosonic ghosts.

A closer look at the topological symmetries (\ref{eetopofs}) and our gauge-fixing conditions shows that this is not the case.  In the relativistic topological sigma models, the holomorphicity conditions fix the gauge of the topological symmetry (\ref{eetopofs}) almost perfectly, leaving only a finite-dimensional space of moduli.  On the other hand, in the tropicalized case, our localization equations fix the topological symmetries (\ref{eetopofs}) less perfectly, leaving an infinite-dimensional space of classical solutions, with the residual $\alpha$ symmetries.  Indeed, recall that on a simply-connected neighborhood $\CU$, in the local coordinates, we found that the classical solutions of the localization equations,
\bea
X&=&X_0(r),\\
\Theta&=&\Theta_0(r)+\theta\p_rX_0(r),
\eea
contain arbitrary functions of $r$, which are precisely acted on by the residual, unfixed $\alpha$ symmetry.

Similarly, the remaining auxiliary field $\beta$ exhibits its own $\alpha$-type symmetry:  The action is invariant under $r$-dependent constant shifts of $\beta$,
\be
\delta\beta=\alpha_\CB(r), 
\ee
with all other fields invariant.  The equations of motion for $\beta$ are solved by
\be
\beta=\beta_0(r).
\ee

While it is not inconceivable to have a path-integral localization to an infinite-dimensional space of solutions, this is not the strategy we wish to pursue here -- our intention is to construct a tropical version of topological sigma models which is dependent only on the Jordan structure, but otherwise it is as close as possible to the standard relativistic topological sigma models.  Therefore, in a way which will become clear below, we will aim to gauge-fix the residual $\alpha$ gauge symmetry in a simple way, simply by using them to set locally $X_0(r)$, $\Theta_0(r)$ and $\beta_0(r)$ to zero (or some other convenient fixed value).  This step will again leave behind only a finite-dimensional moduli space of solutions, and without any need for any secondary ghosts, due to the simplicity of this additional gauge-fixing condition for $\alpha$.

When we ask the same questions on a sleeve $\Sigma$, the $\alpha$ gauge transformation will now have to respect the periodicities of $\theta$ and $\Theta$.  That reduces the gauge transformation parameter $\alpha_1(r)$ to be a constant $\alpha_{1,0}$.  This is in turn consistent with the fact that the classical solutions of the localization equations on the sleeve take the form
\bea
X&=&x_0+nr,\qquad n\in\MZ,\\
\Theta&=&\Theta_0(r)+n\theta,\\
\beta&=&\beta_0(r).
\eea
$\Theta_0(r)$ and $\beta_0(r)$ can be set to zero (or other convenient fixed values) using the $\alpha_0(r)$ and $\alpha_\CB(r)$ gauge symmetries, and $x_0$ can be set to zero (or another convenient constant value) by the constant residual gauge transformation $\alpha_{1,0}$.  The integer winding number $n$ is a gauge invariant under all $\alpha$ gauge transformations.

\subsection{Observables}

Having clarified that our BRST multiplets do not lead to ghost-for-ghosts, it is now clear that the BRST invariant observables of the theory are exactly as in standard, relativistic topological sigma models.  With any differential form on $M$,
\be
\omega_{i_1\ldots i_p}dY^{i_1}\wedge\ldots\wedge dY^{i_p}
\ee
we associate a local operator
\be
\CO^{(0)}_\omega=\omega_{i_1\ldots i_p}\psi^{i_1}\ldots \psi^{i_p}.
\ee
Since no secondary ghosts-for-ghosts were required by the $\alpha$ symmetries, $\psi$ is BRST invariant, therefore $\CO^{(0)}_\omega$ is BRST invariant if $\omega$ is closed, and BRST trivial if $\omega$ is exact.

The Stora-Zumino descent equations then yield a hierarchy of 1-form and 2-form observables on $\Sigma$,
\bea
d\CO^{(0)}_\omega&=&\{Q,\CO^{(1)}],\\
d\CO^{(1)}_\omega&=&\{Q,\CO^{(2)}],\\
d\CO^{(2)}_\omega&=&0.
\eea
(Here as usual the symbol ``$\{Q,\ ]$'' denotes a commutator or anticommutator, depending on the statistics of the object in its second entry.)


\section{Example: The tropical $\MC P^1$ model}

In the case of relativistic topological sigma models, $\MC P^1$ was used as an early test example in \cite{wittentg}  to probe the viability and self-consistency of the BRST construction of the path integral.  In our case, the tropicalization of this example precisely corresponds to keeping just one pair $(X,\Theta)$ in our tropological sigma model, and thus this example enjoys a more privileged status than it did in the relativistic case:  it represents the basic building block from which tropicalizations of higher-dimensional manifolds are naturally built.

\subsection{Observables and correlation functions at tree level}

The fundamental observables of the $\MT P^1$ theory are given by the BRST cohomology of local operators; just as in the relativistic $\MC P^1$ case \cite{wittentg}, this space is two-dimensional, spanned by the identity operator and the operator of ghost number two associated with the top-degree cohomology class $[\omega]$ on $\MC P^1$, which we will simply refer to as $\CO_\omega$.  Consider first the relativistic case: We will choose a representative two-form $\omega$ in the cohomology class $[\omega]$, and normalize it such that
\be
\int_{\MC P^1}\omega=1,
\ee
in which case
\be
\int_\Sigma \Phi^\ast(\omega)=k\in\MZ
\ee
is the instanton number of the map $\Phi$, corresponding to the topological winding number of maps $S^1\to S^1$.  The most natural choice of $\omega$ is the covariantly constant volume element on $\MC P^1$ viewed as a round $S^2$.

In the tropical case, while the BRST cohomology that we have found is identical to that of the relativistic case, we must be a bit more careful with the choice of a representative of $[\omega]$:  We cannot simply choose $\omega$ to be $dX\wedge d\Theta$ multiplied by a suitable normalization constant, since the standard integral $\int_{\MT P^1}dX\wedge d\Theta$ is infinite.  The natural choice is to pick a two-form which is translationally invariant along $\Theta$ but not along $X$, such that the integral converges and can be normalized to 1.  The simplest natural choice is to pick a reference point $X_0\in \MR$, and to choose
\be
\omega_\MT= \frac{1}{2\pi}\delta(X-X_0)dX\wedge d\Theta.
\ee
This representative is then correctly normalized, giving
\be
\int_{\MT P^1}\omega_\MT =1,
\ee
where we interpret the integral of a two-form over $\MT P^1$ in the classical sense of integrals of classical two-forms over the underlying topological space $S^2$.  The choice of the reference point $X_0$ is immaterial, as changing it is cohomologically trivial and thus preserves the cohomology class $[\omega]$.  Note that the role of the instanton number $k$ in the tropical case, 
\be
\int_\Sigma \Phi^\ast(\omega_\MT)=\frac{1}{2\pi}\int_0^{2\pi}\frac{\p\Theta}{\p\theta}d\theta=k\in\MZ,
\label{eetropinstk}
\ee
is played by the winding number of the worldsheet periodic coordinate $\theta$ around the periodic target space dimension $\Theta$. 

It is intriguing that one can take an alternative (but essentially equivalent) perspective on how to choose and interpret the representative of $[\omega]$ in the tropical case:  We can insist that $[\omega]$ be represented by the translationally invariant form
\be
\omega'_\MT=\frac{1}{2\pi}dX\wedge d\Theta,
\ee
if we simultaneously change the interpretation of what it means to integrate this form over $\MT P^1$ (and of its pull-back via $\Phi$ over $\Sigma$) -- the integral must now be interpreted in the tropical sense \cite{msintro,rau,mikhalkinrau} along the tropicalized dimension $X$,
\be
\int_{\MT P^1}^\oplus\omega'_\MT\equiv \frac{1}{2\pi}\int^\oplus dX\int_0^{2\pi}d\Theta =1.
\label{eetropint}
\ee
Indeed, the tropical version of the integral of a differential form $f(X)dX$ along the one-dimensional tropical variety parametrized by $X$ is simply defined by finding the maximum value of $f(X)$ over the range $\CR$ of integration,
\be
\int^\oplus_\CR f(X)dX\equiv\textrm{max}\{f(X), X\in\CR\}.
\ee
Since our form has $f(X)=1$, a constant, our result (\ref{eetropint}) follows from the definition of the tropical integration.  Similarly, the instanton number $k$ would result from the tropical integration along the worldsheet dimension $r$ of the pull-back $\Phi^\ast(\omega'_\MT)$.  

Now we wish to calculate the correlation functions of the local observables, first on the worldsheet with genus zero.  Let us first recall how the evaluation of the correlation functions of $\CO_\omega$'s proceeds in the standard relativistic case with the $\MC P^1$ target \cite{wittentg}.  Instantons for worldsheets of genus zero are rational complex curves, described by
\be
Z(z)=a\frac{(z-b_1)\ldots(z-b_k)}{(z-c_1)\ldots(z-c_k)}.
\ee
Here $Z\in\MC$ is the standard coordinate on $\MC P^1$ with the point at infinity removed, and $z$ is the standard worldsheet coordinate on $\Sigma=S^2$ with the point at infinity removed.  The integer $k$ is the instanton number, and the moduli space of instantons with the instanton number $k$ is complex $2k+1$-dimensional, parametrized by the complex parameters $a,b_i,c_i$, with $i=1,\ldots k$.  We will introduce a coupling $\lambda$, designed such that the contribution from the instantons of instanton number $k$ to any correlation function is weighted by a factor of
\be
\lambda^k.
\ee
The generic correlation function will then vanish (for example, due to the existence of unsaturated ghost zero modes), unless we select $2k+1$ points $P_0,P_1,\ldots P_{2k}$ on $\Sigma$ and calculate the correlation function of $2k+1$ observables $\CO_\omega$ inserted at these points.  Each insertion of $\CO_\omega(P_\ell)$ is equivalent to the restriction of the value of $Z(z)$ at $z=P_\ell$ to equal a fixed prescribed value $Z_\ell$; the simplest and most convenient choice for these values is to take $Z_\ell=0$ for $\ell=1,\ldots, k$, $Z_\ell=\infty$ for $\ell=k+1,\ldots,2k$, and finally $Z_0=1$.  This leaves precisely one rational function of instanton number $k$, with the prescribed structure of zeros and poles and with the fixed overall normalization.  The correlation function at genus $g=0$ thus is
\be
\left\langle\CO_\omega(P_0)\CO_\omega(P_1)\ldots \CO_\omega(P_{n})\right\rangle_{g=0}=
\left\{
\begin{array}{l}0,\qquad\ \ \textrm{if}\ n=2k-1;\cr \lambda^k,\qquad \textrm{if}\ n=2k.\cr\end{array}\right.
\label{eeocorr}
\ee
(The BRST cohomology of point-like operators in this model of course contains also the operator $\CO_1(P)$ that corresponds to the zero-form constant cohomology class $[1]$; but its effect in the path integral is trivial, in the sense that it does not restrict the instanton at point $P$ in any way, and therefore can be inserted any number of times inside the correlation functions without changing their values as given in (\ref{eeocorr}).)

In the tropical case, we also wish to evaluate the family of correlation functions of any number of $\CO_\omega$'s, inserted at points $P_\ell$ of the worldsheet.  As we discussed above, the solutions of our localization equations (which one could refer to as ``tropical instantons'') are also assigned an instanton number $k$, now defined via (\ref{eetropinstk}), or equivalently with the use of tropical integration (\ref{eetropint}).  For virtually identical reasons as in the relativistic case, for such an instanton of instanton number $k$ to give a nonzero contribution to the correlation function, we need an insertion of $2k+1$ judiciously placed observables $\CO_\omega$ so that the instanton is an isolated solution without nontrivial moduli.

While we will see that the result for the correlation functions in the tropical case reassuringly coincides with the relativistic result (\ref{eeocorr}), the steps of the evaluation in the tropological sigma model are intriguingly different from the standard relativistic theory.  Our instantons will be globally well-defined solutions of our localization equations, mapping $\Sigma$ (which is topologically a sphere, with a number of marked points where the observables are inserted), to the $\MT P^1$ target space.  As a starting point, we need to choose a fixed, generic Jordan structure on $\Sigma=S^2$, capable of supporting an instanton of instanton number $k$.  In order for such solutions to exist and to be isolated with no moduli, a certain number of insertions of $\CO_\omega$ is again needed.  It is easy to see that for a generic Jordan structure, the instanton of instanton number $k$ again requires the insertion of precisely $2k+1$ operators $\CO_\omega(P_\ell)$, just as in the relativistic case.  We wish to mimic the relativistic evaluation as possible, to stress the parallels and to see the differences in the construction.  Thus, following the relativistic example, in order to get a unique instanton with no moduli we will require points $P_1,\ldots P_k$ to be mapped to the tropical zero, $X=-\infty$; similarly, points $P_{k+1},\ldots P_{2k}$ will be mapped to the tropical infinity, $X=+\infty$; and the remaining point $P_0$ will be mapped to any finite point in the target, say $X=0$, $\Theta=0$.

\begin{figure}[t!]
  \centering
    \includegraphics[width=0.45\textwidth]{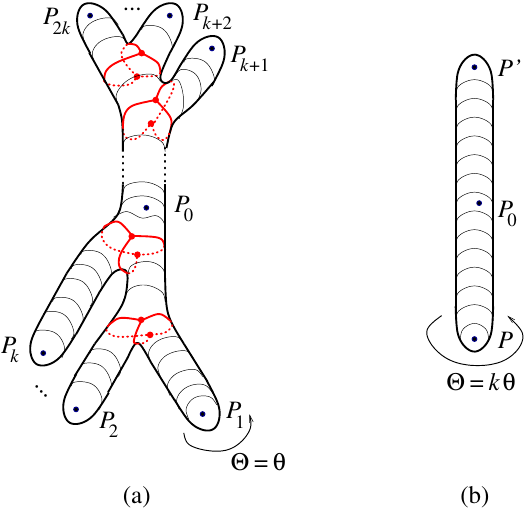}
    \caption{The instanton contributing to the $2k+1$-point function. \textbf{(a):} A map from the worldsheet with a generic Jordan structure.  \textbf{(b):} The same instanton on a worldsheet with a non-generic Jordan structure.}
    \label{ffinstant}
\end{figure}

The generic Jordan structure on $\Sigma$ for which such an isolated tropological instanton solution of instanton number $k$ exists is depicted in Figure~\ref{ffinstant}(a):  It is given by a foliated surface with $2k$ punctures $P_1,\ldots P_{2k}$ and one additional marked point $P_0$ on one of the nonsingular leaves of the foliation.  Moreover, the $2k$ punctures (which represent semi-infinite sleeves) are naturally divided into two groups of $k$, one group considered ``incoming'' where the natural value of the adapted coordinate $r=-\infty$ and the other group ``outgoing'', with $r=+\infty$.  Generically, such a surface contains junctions of three sleeves at a time.  Consecutive singular leaves are connected by finite sleeves, whose lengths represent moduli of the Jordan structure; since gravity is still non-dynamical, these moduli are fixed numbers, chosen to have generic nonzero values.

The unique instanton solution on this surface is constructed as follows.  First, the placement of $\CO_\omega$ at $P_1,\ldots, P_k$ constrains the values of the instanton solution to be
\be
X(r=-\infty)=-\infty,
\ee
with $\Theta$ on the individual half-infinite sleeve around each of these $k$ punctures exhibiting the winding number $n=1$.  Similarly, the insertions of $\CO_\omega$ at the ``outgoing'' punctures at $P_{k+1},\ldots, P_{2k}$ constrain the values
\be
X(r=\infty)=+\infty,
\ee
for each of these half-infinite sleeves, again with $\Theta$ exhibiting winding number one around each such sleeve.  The rest of the instanton is simply constructed by the matching construction at singular leaves where three sleeves meet, as presented in \S\ref{ssadmis}.  The winding numbers are conserved at each such junction, assuring that the map from $\Sigma$ to the target space is globally well-defined (and in fact smooth).  A quick check shows that such a map exhibits the desired instanton number $k$, defined simply by the worldsheet integral of the pull-back of the two-form $\omega$, either in the conventional or the tropical sense as discussed at the beginning of this subsection.

Is such an instanton solution unique?  Not quite yet; as we construct the global surface map by starting at points with $r=-\infty$ on the worldsheet and following the globally defined function $r$ on $\Sigma$ towards $r=+\infty$, we encounter singular sleeves where two incoming sleeves form one outgoing sleeve, or one incoming sleeve splits into two outgoing ones.  Using the $\alpha$ gauge transformations to match the values of $X,\Theta$ of the sleeves that meet at a singular leaf, the continuity of $X(r)$ can be maintained throughout the entire construction of the global map, but we are still left with one undetermined zero mode:  This instanton still has one modulus $X_0$, corresponding to the unfixed freedom to shift the target space coordinate by a constant,
\be
X(r)\to X(r)+X_0,
\ee
globally everywhere throughout the surface $\Sigma$.  This remaining modulus will then be eliminated simply by choosing our bulk point $P_0$ on some generic nonsingular leaf of the foliation, and requiring
\be
X(P_0)=0,  
\ee
yielding a unique isolated instanton.  Thus, the correlation functions in the tropological $\MT P^1$ model at string tree level match exactly the relativistic result, (\ref{eeocorr}).

What if we choose a different Jordan structure, with a different number of incoming and outgoing punctures?  We claim that the standard BRST arguments (whereby the deformation of the Jordan structure can be viewed at least formally as a BRST-exact operation) suggest the correlation functions will not change.  This can be tested and further supported by considering various examples.  Consider for example the Jordan structure depicted in Figure~\ref{ffinstant}(b), with only one incoming and one outgoing puncture.  Again, a unique instanton of instanton number $k$ can be constructed:  Now we are forced to put all the $\CO_\omega(P_1),\ldots,\CO_\omega(P_k)$ at the unique incoming puncture, so we must set
\be
P_1=\ldots= P_k=P.
\ee
The winding number around the unique sleeve must therefore be $k$.  Similarly, all $\CO_\omega(P_{k+1}),\ldots,$ $\CO_\omega(P_{2k})$ must be placed at the unique outgoing puncture, 
\be
P_{k+1}=\ldots= P_{2k}=P'
\ee
(otherwise no solution would exist).  As on the generic surface from Figure~\ref{ffinstant}(a), we must fix the overall zero-mode of $X_0$ the instanton by setting $X(P_o)=0$, thus leading to the unique nontrivial instanton solution and hence the same correlation functions that we first obtained with the generic choice of the Jordan structure.

Similarly, if you start with a Jordan structure on $S^2$ which has more than $2k$ punctures, any solution of the localization equations with instanton number $k$ would have some of these punctures describing sleeves whose winding number must be zero.  Such sleeves can then be collapsed to a point, and since they do not restrict the value of the instanton at that point, they correspond to the harmless insertion of the $\CO_1$ operator which as we know does not change the value of the correlation functions.

\subsection{Correlation functions to all loop orders}

Having completed the evaluation of correlation functions of BRST invariant point-like operators at genus zero, it is natural to extend the analysis to all string loop orders.  In the relativistic case, the structure of topological theories without gravity can be usefully encoded in the axiomatic approach initiated by Atiyah \cite{atiyah} and Segal \cite{segal}.  In the case of our tropological sigma models, one can similarly anticipate that an appropriate refinement of the Atiyah-Segal axioms will be valid:  Our theory turns out to be defined by a more-or-less standard path integral (albeit nonrelativistic), and the theory is not yet coupled to dynamical gravity; therefore, it should exhibit similar axiomatic properties as the relativistic theories before coupling to gravity.

In this section, we will not assume the validity of such axioms, and instead derive the desired all-loop correlation functions for our $\MT P^1$ example by direct arguments.  The main tool needed to extend our tree-level correlation functions to higher orders is the inclusion of a handle into the foliated worldsheet.  Such a handle should be equipped with its own consistent foliation, such that there exists a unique nontrivial instanton at the corresponding instanton number, contributing to the correlation function as in the tree-level case.  In analogy with the relativistic case \cite{wittentg}, one can then expect that the addition of such a handle would correspond to the insertion of a local ``handle operator'' $\CW$.

\begin{figure}[t!]
  \centering
    \includegraphics[width=0.3\textwidth]{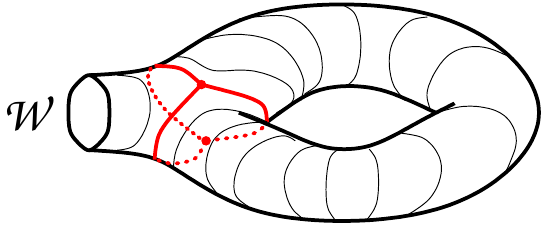}
    \caption{The handle operator $\CW$ whose insertion raises the genus of $\Sigma$ by one.}
    \label{fftorus}
\end{figure}

It appears that the only such foliated handle represented by a nontrivial $\CW$ operator is the torus with one finite sleeve attached, and equipped with a generic foliation structure, as depicted in Figure~\ref{fftorus}.  The instanton solution on such a handle can be constructed as follows.  The finite sleeve that closes onto itself to form the torus will support a general solution, specified by the winding number $n$.  But to ensure that the loop can be closed in the direction transverse to the foliation requires that $X(r)$ be periodic along that closed loop, forcing $n=0$.  Thus, the entire loop formed by the closed finite sleeve is mapped to a single point in $\MT P^1$.  Similarly, by the winding-number conservation at the singular sleeve of the handle in Figure~\ref{fftorus}, the winding number of the finite sleeve attached to the torus must vanish, and the sleeve can also be collapsed to a point.  Thus, the entire handle will be represented on the rest of the worldsheet by an insertion of a local operator $\CW$, without any need to invoke a factorization property of the path integral.  As in the relativistic case, this local operator must be given by
\be
\CW=2\CO_\omega.
\ee
(This follows simply from the matching to the one-loop partition function, interpreted as the torus one-point function of $\CO_1$.  Since the two-point function $\langle\CO_\omega\CO_1\rangle_0=1$, this means that $\CW$ is proportional to $\CO_\omega$.  Since the same one-loop partition function can also be interpreted as the trace over the two-dimensional space of physical states, the overall normalization must be equal to two, just as in the relativistic case.  Finally, the handle operator must be proportional to $\lambda^0$, since the winding number through the handle is zero, and therefore adding the handle does not change the value of the instanton number.)

With this handle operator $\CW$ at hand, we can add additional handles, as long as the singular leaves of the worldsheet foliation at which these handles are inserted are generic with respect to each other.  This again reproduces the relativistic result for higher-genus correlation functions.  In particular, the partition function -- summed to all orders in the string coupling $g_s$ -- gives 
\be
\left\langle\!\left\langle\CO_1\right\rangle\!\right\rangle
\equiv \sum_{g=0}^\infty g_s^{2g-2}\langle\CO_1\rangle_g=
\sum_{g=0}^\infty g_s^{2g-2}\langle\CO_1\CW^{g-1}\rangle_0=\frac{2}{1-4g_s^4\lambda},
\ee
again matching the well-known relativistic result of the $\MC P^1$ model \cite{wittentg}.

Should there be nontrivial handle operators associated with more complicated ways how to insert topological handles on foliated surfaces?  Consider for example inserting a handle by replacing a finite sleeve in a tree-level diagram with a torus with two finite sleeves attached -- a foliated surface with two boundaries that would follow from gluing the two surfaces depicted in Figures~\ref{ffwsjordan}(c) and \ref{ffwsjordan}(d) along two of their boundary components.  A generic Jordan structure on this surface would contain the lengths of the two internal sleeves inside the loops as fixed moduli.  In the generic case, the ratio of these lengths is a generic real number, generally not a rational number.  But for the solutions along the sleeves to have the correct periodicity along $\Theta$, $X(r)$ along the two sleeves must be fixed to be $n_1r$ and $n_2r$, with $n_1$ and $n_2$ both integers.  The condition of periodicity of $X$ as the loop is getting closed would then require $n_1/n_2$ to be equal to the generic real ratio of the two lengths of the sleeves, and therefore is not possible to satisfy if the Jordan structure is indeed generic, and if $n_1$ and $n_2$ are both nonzero.  If, on the other hand, one of the $n_1$ and $n_2$ is zero, the corresponding sleeve can be contracted to a point, reducing the construction to the $\CW$ operator discussed above. With $n_1$ and $n_2$ both nonzero, there is no instanton supported by this handle, and no need to invent a corresponding representation of such a handle by local operators.  We are satisfied to see that the correlation functions of the tropological sigma model, to all string loops, appear to have just enough content to reproduce precisely the results for its relativistic cousin.

\section{Continuations to real worldsheet time}

In the relativistic case, the topological sigma models represent much simpler and controllable cousins of the physical theories with propagating degrees of freedom, which could be either the untwisted supersymmetric sigma models, or the sigma models truncated to their bosonic sector.  One can naturally ask whether a similar relationship exists between the nonrelativistic tropological sigma models, and theories with propagating local degrees of freedom.

Consider the bosonic sector of our tropological sigma model, with action
\be
S=\int dr\,d\theta\left\{\frac{1}{2}\left(\p_\theta\Theta-\p_rX\right)^2+\beta\p_\theta X\right\}.
\label{eeactph}
\ee
The invariance of the action under constant translations in $r$ and $\theta$ and the Noether theorem imply the existence of a conserved energy-momentum tensor $T^\alpha{}_\beta$,
\bea
T^r{}_r&=&\frac{1}{2}(\p_rX)^2-\frac{1}{2}(\p_\theta \Theta)^2-\beta\p_\theta X,\\
T^r{}_\theta &=&(\p_r X-\p_\theta\Theta )\p_\theta X,\\
T^\theta{}_r&=&-(\p_r X-\p_\theta\Theta )\p_r\Theta+\beta\p_r X,\\
T^\theta{}_\theta &=&\frac{1}{2}(\p_\theta\Theta)^2-\frac{1}{2}(\p_r X)^2.
\eea
Note some similarities and a few slight conceptual differences compared to the relativistic case: This energy-momentum tensor is conserved, $\p_\alpha T^\alpha{}_\beta=0$, but because of the absence of any non-degenerate metric on $\Sigma$, it is not natural to raise or lower the indices on $T^\alpha{}_\beta$.  The lower index indicates which translation symmetry this conserved current is associated with, while the upper index labels the components of this current.  (One could use $g_{\alpha\beta}$ and $h^{\alpha\beta}$ to lower or raise indices, but after this operation, some information contained in $T^\alpha{}_\beta$ would be inevitably lost.)  Note that the trace of the energy-momentum is still well-defined,
\be
T^\alpha{}_\alpha=-\beta\p_\theta X.
\ee
It vanishes on-shell, confirming that our theory exhibits isotropic scaling with dynamical exponent $z=1$.  Note that despite the isotropic scaling, the theory is still sensitive to the worldsheet foliation structure, and the symmetries generated by the energy momentum tensor are indeed in accord with its underlying nonrelativistic conformal symmetry.

\subsection{Looking for real time in the cross-channel: Conformal symmetries}

We would like to understand whether this theory can be appropriately formulated in real worldsheet time, with propagating degrees of freedom, not just as a bosonic sector of a topological theory.  For reasons we explain below, we find it more interesting to consider the real worldsheet time to run \textit{along the leaves of the foliation}, and not in the direction transverse to the foliation (as was the case for the imaginary time in the tropological sigma models).%
\footnote{Such theories can then be coupled to a suitable version of nonrelativistic gravity, which would correspond in this channel to the $c\to 0$ nonrelativistic limit, often referred to in the literature as the Carrollian limit.}
We would like to identify a real-time interpretation of the theory described by (\ref{eeactph}), perhaps by a suitable analytic continuation of the fields and coordinates, such that it satisfies sensible physical requirements, including unitarity, and positivity of energy.

\begin{figure}[t!]
  \centering
    \includegraphics[width=0.3\textwidth]{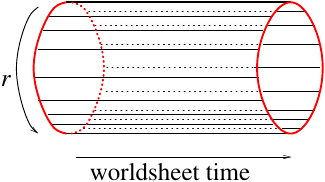}
    \caption{The worldsheet foliation in the cross-channel.}
    \label{ffcross}
\end{figure}

A naive Wick rotation $\theta\to it$ to candidate real time $t$ would not work:  $S$ as given in (\ref{eeactph}) is real (for real fields, and real coordinates $r,\theta$).  A continuation $\theta\to it$ would make the action complex-valued for real $t$.  In real-time physical theories, one can ensure unitarity by requiring the real-time action to be real.  When such a theory is Wick-rotated from real to purely imaginary time, the original condition of unitarity is reformulated as ``reflection-positivity'':  The imaginary-time action can have a non-zero imaginary part, which is constrained to be odd under the time reversal transformation.  We will impose the same restrictions on the expected behavior of $S$ in our nonrelativistic case as well.

Now that we have decided that we would like to interpret the direction along the leaves of the foliation as time, it is natural to compactify $r$ on a circle.  In addition, we will choose the coordinate $r$ such that it is periodic with periodicity $2\pi$.  In this channel, we can study in some more detail the worldsheet conformal symmetries generated by $f(r)$ and $F(r)$, as found above in the tropological context:  Unlike the case of noncompact $r$, there is now a natural countable basis that spans the Lie algebra of conformal symmetries, obtained simply by decomposing the infinitesimal generators $f(r)$ and $F(r)$ into Fourier modes,
\be
f(r)=\sum_\MZ L_me^{imr},\qquad F(r)=\sum_\MZ J_me^{imr}.
\ee
At the classical level, they satisfy a Lie algebra with a clear geometric interpretation:  $L_m$ form the Virasoro algebra of the infinitesimal diffeomorphisms of $r\in\BS^1$, while $J_m$ represents an Abelian algebra, transforming under the Virasoro as the Fourier components of a spin-two field.  Quantum mechanically, this algebra admits two central charges $c$ and $c_\times$,
\bea
[L_m,L_n]&=&(m-n)L_{m+n}+\frac{c}{12}m(m^2-1)\delta_{m+n,0},\cr
[J_m,J_n]&=&0, \cr
[L_m,J_n]&=&(m-n)J_{m+n}+\frac{c_\times}{12}m(m^2-1)\delta_{m+n,0}.
\eea
Note that the Abelian commutation relations between $J_m$'s do not allow a nonzero central extension consistent with the commutation relations involving the $L_m$ generators (see also \cite{compere}).

This algebra has been recognized and studied in the literature, in several contexts \cite{b,bi,bii,biii}.  It is also recognized as the Bondi-Metzner-Sachs (BMS) algebra of asymptotic symmetries of the Minkowski spacetime at infinity \cite{sachso,bondi,sachs,sachsgw,bicak}, in the special case of the $2+1$ dimensions, and it has been studied extensively in that context \cite{h,db,biv,bv,hen,bvi,bvii,bvii}.

\subsection{Quantizing with $\theta$ as the real time}

Since $S$ in (\ref{eeactph}) is already real, we can try to interpret $\theta$ itself as \textit{real} time, and see whether its canonical quantization (using methods of Dirac quantization \cite{htbook}) gives consistent results.  Interpreting now $\theta$ as time, the canonical momenta $P,\Pi$ and $\pi$ conjugate to $X,\Theta$ and $\beta$ are
\bea
P&=&\beta,\\
\Pi&=&\p_\theta\Theta-\p_rX,\\
\pi&=&0.
\eea
The system can be quantized straightforwardly.  The vanishing of $\pi$ is a constraint, which is very easy to deal with: $X$ and $\beta$ simply represent an unconstrained canonical pair.  The Hamiltonian density and the Hamiltonian are 
\be
\CH=\frac{1}{2}\Pi^2+\Pi\p_rX,\qquad H=\int dr\,\CH.
\ee
This is of course a Hamiltonian of some free theory.  The individual degrees of freedom can be identified by performing a canonical transformation that diagonalizes the Hamiltonian.  This canonical transformation is given by
\bea
\tilde\Pi&=&\Pi+\p_rX,\qquad \,\tilde X=X,\\
\tilde P&=&P-\p_r\Theta ,\qquad\ \tilde\Theta=\Theta.
\eea
(Here the first line is suggested by completing the square in the Hamiltonian, and the second is required to make the full transformation canonical, since the new momentum $\tilde\Pi$ does not commute with the old $P$.)  After the canonical transformation, the Hamiltonian density becomes
\be
\CH=\frac{1}{2}\tilde\Pi^2-\frac{1}{2}(\p_r\tilde X)^2,
\label{eehamtilde}
\ee
and we see a problem:  This theory with $\theta$ interpreted as real time has a real action and is formally unitary, but its energy is not bounded from above or below.  

\subsection{Conformal reductions}

Having found that the system represents two decoupled free degrees of freedom, one with a positive-definite energy and the other with a negative-definite energy, it is natural to ask whether we can consistently truncate the system (without violating conformal invariance) to keep only one or the other subsystem.

This should not be done in an \textit{ad hoc} manner, but more systematically -- perhaps by invoking a suitable gauge symmetry.  We have found that our theory has a subtle gauge invariance represented by the $\alpha$ symmetries, which depend arbitrarily on $r$ but only linearly on $\theta$.  In the Lagrangian formalism, it is not quite clear whether or how such hybrid symmetries can be used to reduce the number of degrees of freedom from two to one.  We will therefore fall back on the canonical quantization in the Hamiltonian formalism for systems with constraints.

The decoupling of the two subsystems is manifest in the $\tilde X,\tilde P,\tilde\Theta, \tilde\Pi$ coordinates on the phase space.  We first attempt to eliminate the $\tilde X,\tilde P$ canonical pair.  There are two paths how to do so, leading to the same result:  First, we choose to impose
\be
\tilde X=0
\ee
as a primary constraint.  Since the constraint commutes with the Hamiltonian, there is no secondary constraint.  Our constraint generates a gauge symmetry, acting via the commutator with the constraint.  In particular, under this gauge symmetry, with gauge parametrer $\tilde\alpha_1$, $\tilde P$ transforms simply via
\be
\delta P=\tilde\alpha_1.
\ee
Setting $\tilde P=0$ is then a good gauge choice, yielding the desired reduction to just one degree of freedom, which can be described by the reduced action
\be
S_1=\int dr\,d\theta\,\frac{1}{2}(\p_\theta\Theta)^2.
\label{eeredone}
\ee
The second path leading to the same result starts instead by imposing the primary constraint
\be
\tilde P=0.
\ee
The commutator with the Hamiltonian reveals that there is now a secondary constraint,
$\tilde X=0$, and no gauge symmetries.  The constraints form a second-class pair, which when treated via the Dirac bracket simply eliminates the $\tilde X,\tilde P$ canonical pair, leading to the same reduced theory described by (\ref{eeredone}).

If we instead wish to eliminate the $\tilde\Theta,\tilde\Pi$ canonical pair, we again have two paths:  We can choose the primary constraint
\be
\tilde\Theta=0,
\ee
which will imply a secondary constraint $\tilde\Pi$, forming a second-class pair; the Dirac bracket quantization of the system then eliminates the $\tilde\Theta,\tilde\Pi$ pair, leading to one degree of freedom described by the reduced action
\be
S_2=\int dr\,d\theta\left\{\frac{1}{2}(\p_r X)^2+\beta\p_\theta X\right\}.
\label{eeredtwo}
\ee
(Now of course the energy of the system would be negative-definite, which we can simply compensate for by reversing the overall sign in front of the original full action and repeating the reduction steps.)

The second path starts with imposing
\be
\tilde\Pi=0
\ee
as our primary constraint.  There is no secondary constraint.  The primary constraint is again first-class, generating a gauge symmetry with gauge parameter $\tilde\alpha_2$, acting as
\be
\delta\tilde\Theta=\tilde\alpha_2.
\ee
Setting $\Pi=0$ is a good gauge-fixing condition, reducing the system again to (\ref{eeredtwo}).

Note that the action of the gauge symmetries $\tilde\alpha_1$ and $\tilde\alpha_2$ in the Hamiltonian quantization is quite reminiscent of the $\alpha$ gauge symmetries that we found in the Lagrangian formalism.  How such symmetries should be correctly implemented in the Lagrangian quantization is somewhat obscure, since they are a hybrid between a local and a global symmetry; however, the Hamiltonian quantization with constraints has provided a natural guidance how to interpret and implement such symmetries in the bosonic theory.

Are the two truncations to a system with just one degree of freedom consistent with conformal invariance?  A closer look at the conformal transformation rules for all fields reveals a happy fact:  The same split into the $\tilde X,\tilde P$ and $\tilde\Theta,\tilde\Pi$ pairs that resulted from our canonical transformation in the Hamiltonian framework would also result if we instead required that the split is conformally invariant in the Lagrangian formalism.  Thus, our Hamiltonian quantization has produced two consistent conformal truncations of the original system, to subsystems with Lagrangians (\ref{eeredone}) and (\ref{eeredtwo}).

\subsection{Another analytic continuation to real time}

Having seen that the original system (\ref{eeactph}) has two consistent conformal truncations, one can naturally wonder whether one can consistently analytically continue the theory so that (i) both $\tilde X$ and $\tilde\Theta$ degrees of freedom are kept, (ii) the theory has a suitably continued conformal symmetry, and (iii) it is unitary, with energy bounded from below.  If such a continuation exists, it would also be interesting to see whether it comes from a suitably continued theory in imaginary time, which satisfies the reflection positivity condition on the action.

In the original variables $X,\Theta,\beta$ appearing in (\ref{eeactph}), it is not immediately obvious how to perform such a desired continuation.  Wick rotations of $r$ or $\theta$ and analytic continuations to some of the fields from real to purely imaginary values do not work.  In order to find the correct analytic continuation, let us first compare the theory before and after the canonical transformation to $\tilde X,\tilde P,\tilde \Theta, \tilde\Pi$ variables, and ask how this transformation manifests itself in the Lagrangian formalism.  From the known Hamiltonian (\ref{eehamtilde}), we can reverse-engineer the Lagrangian and express it in the original Lagrangian variables,
\bea
\tilde L&=&\tilde P\p_\theta \tilde X+\tilde\Pi\p_\theta\tilde\Theta-\CH\\
&&\qquad\qquad{}=\frac{1}{2}(\p_\theta\Theta)^2+\frac{1}{2}(\p_r X)^2+(\beta-\p_r\Theta)\p_\theta X.
\eea
This Lagrangian differs from the original Lagrangian in (\ref{eeactph}) simply by adding what is locally a total-derivative term; more precisely, the actions differ globally a topological invariant, given by the integral of the pull-back of $d\Theta\wedge d X$ to $\Sigma$:
\be
L=\tilde L+\p_r\Theta\p_\theta X-\p_\theta\Theta\p_r X.
\label{eelltil}
\ee

Perhaps more importantly, in this representation we see a hint how to perform the desired analytic continuation to real time, with positive-definite energy and real action:  We need to Wick rotate $\theta$ to real time $t$ while keeping $X$ and $\Theta$ real, and then we must rotate the shifted field
\be
\beta'\equiv \beta-\p_r\Theta
\ee
from real to purely imaginary values, $\beta'=i\SB$, with $\SB$ real.

Is such an analytic continuation consistent with the appropriately analytically continued conformal symmetries?  After the Wick rotation $\theta=it$, we must make sure that the conformal transformations maintain the reality of $r$ and $t$.  This is achieved by keeping $f(r)$ real, and rotating $F(r)=i\Phi(r)$ to be purely imaginary.  Such analytically continued conformal transformations act naturally on $X$ and $\Theta$, while preserving their reality.  In addition, on $\beta'$ they act via
\be
\delta\beta'=f(r)\p_r\beta'+\beta'\p_rf(r)+\left(\vphantom{\frac{1}{2}}\Phi(r)+t\p_r f(r)\right)\p_t\beta'-i\p_r X\p_r\left(\vphantom{\frac{1}{2}}\Phi(r)+t\p_r f(r)\right).
\ee
This is indeed consistent with analytically continuing $\beta'=i\SB$.  The conformal transformations with parameters $f(r)$ and $\Phi$ transform $\SB$ consistently with the reality of $\SB$ while $X$ and $\Theta$ are kept real.  In contrast, note that the original auxiliary field $\beta$ transforms after the Wick rotation in a slightly more complicated way than $\beta'$,
\be
\delta\beta=f(r)\p_r\beta+\beta\p_rf(r)+\left(\vphantom{\frac{1}{2}}\Phi(r)+t\p_r f(r)\right)\p_t\beta+(\p_t\Theta-i\p_r X)\p_r\left(\vphantom{\frac{1}{2}}\Phi(r)+t\p_r f(r)\right).
\ee
The term that contains $\p_t\Theta$ in this transformation represents an obstruction against analytically continuing $\beta$ to purely imaginary values while keeping $X$ and $\Theta$ real -- such a continuation would not be consistent with conformal invariance.  This observation makes it clear why the proper analytic continuation of the original theory (\ref{eeactph}) was rather obscure from the perspective of its original variables.  

Note also that our analytic continuation will make the topological term in (\ref{eelltil}) purely imaginary; if we desire the real-time action to be fully real including the topological term, we need to analytically continue the coupling constant of this topological term (\textit{i.e.}, its ``theta angle'') accordingly.  

\subsection{Motivation from nonequilibrium string perturbation theory}

Our preference to interpret the direction along the leaves of the foliation as the worldsheet time originates from broader considerations in physical string theory, concretely from our expectations about the structure of string perturbation theory when the system it describes is not in equilibrium.  In quantum field theory, systems out of equilibrium are most naturally formulated with the use of the closed, doubled time contour (known as the Schwinger-Keldysh contour), in which the system is first evolved from the far past to the remote future, and then again de-evolved into the remote past.  It is an intriguing open question whether the covariant formulation of string perturbation theory can also be extended to the Schwinger-Keldysh time, and if so, how.

In the absence of a direct worldsheet construction, this question was addressed in \cite{neq,ssk,keq} with the use of expected holographic duality to gauge theories in the large-$N$ expansion, for which the formulation away from equilibrium using the Schwinger-Keldysh contour is known.  The result found in \cite{neq,ssk,keq} strongly suggests that in nonequilibrium string perturbation theory, the standard expression for observable amplitudes in terms of a sum over contributions from worldsheets $\Sigma_g$ of genus $g$, weighted by the appropriate power of the string coupling constant $g_s$, 
\be
\CA=\sum_{g=0}^\infty g_s^{2g-2}\CF(\Sigma_g),
\ee
will be refined in an interested and perhaps unexpected way:  On the Schwinger-Keldysh time contour, each worldsheet $\Sigma_g$ carries a natural triple decomposition,
\be
\Sigma_g=\Sigma^+_{g_+}\cup \Sigma^\wedge_{g_\wedge}\cup\Sigma^-_{g_-},
\label{eestrprt}
\ee
where $\Sigma^+_{g_+}$ corresponds to the part of the worldsheet on the forward branch of the time contour, $\Sigma^-_{g_-}$ corresponds to the backward part of the time contour, and the ``wedge region'' $\Sigma^\wedge _{g_\wedge}$ is the portion of the worldsheet that connects them, and corresponds to the turn-around late time where the forward and backward branch of the time contour meet.  Topologically, the wedge region is connected to the other two parts of this decomposition along a collection of $S^1$ boundaries, but geometrically, there are hints that these connecting boundaries should be interpreted as nodes in $\Sigma_g$.  Of course, the total genus $g$ is uniquely determined in terms of the three genera $g_+$, $g_\wedge$ and $g_-$, such that the Euler number $\chi(\Sigma_g)$ matches the Euler number obtained from the triple decomposition.

The interesting point is that for the matching to the expectations from the large-$N$ expansion on the dual gauge theory side, not only are $\Sigma^+$ and $\Sigma^-$ expected to be of arbitrary genus, but also the wedge region must be allowed to be of any genus $g_\wedge$.  Thus, the perturbative sum (\ref{eestrprt}) is refined into a \textit{triple sum} over nontrivial topologies of $\Sigma^+$, $\Sigma^-$, and also $\Sigma^\wedge$.

Consider now the specific case of critical string or superstring theory.  While the worldsheet path-integral description of the theory in regions $\Sigma^\pm$ might be expected to reproduce the standard Polyakov-like path integral, the connecting $\Sigma^\wedge$ region remains quite mysterious from the point of view of the worldsheet dynamics.  The robust arguments from the large-$N$ duality to nonequilibrium gauge theories are not refined enough to predict clearly a worldsheet description.  However, they do provide one strong hint:  While the topology of $\Sigma^\wedge_{g_\wedge}$ can be arbitrary (i.e., its genus $g_\wedge=0,1,\ldots$), its \textit{geometry} is expected to be highly anisotropic on the worldsheet \cite{ssk}:  The parts of the ribbon diagrams that generate $\Sigma^\wedge$ equip this region with a coarse-grained foliation structure; the leaves of the foliation are segments connecting a point on the boundary with the $\Sigma^+$ region to a point on the boundary with the $\Sigma^-$ region.  Moreover, the worldsheet geometry undergoes an anisotropic limit, such that the lengths along the leaves of the foliation go to zero while the lengths in the transverse directions are held fixed.  This type of anisotropy expected from the holographic duality of the mysterious wedge region is extremely reminiscent of the geometric structure we have seen on the tropicalized worldsheets in this paper, with the coordinate $\theta$ along the foliation leaves now interpreted as the worldsheet time.%
\footnote{Tropical geometry has already made its appearance in the study of equilibrium string amplitudes, first in the $\alpha'\to 0$ limit \cite{tourkine} and more recently at general $\alpha'$ \cite{eberhardti,eberhardtii}.}

\section{Conclusions and outlook}

In this paper, we focused on the tropicalization of the topological A-model, and developed the theory and its symmetries from first principles of the BRST gauge-fixed path integral formulation.  It is natural to ask how the process of tropicalization of the path integral influences various dualities that have been well studied in the relativistic case, such as the relationship with the B-model.  We leave the study of possible generalizations of our construction of A-model tropological sigma models open for future study.  

Even in the case of the A-model, we contented ourselves in this paper with finding the standard point-like BRST invariant observables of the relativistic A-model (the ``quantum cohomology'' of the relativistic target) embedded as observables in our tropological sigma model.  This leaves an open question of whether our nonrelativistic theory could contain additional, more subtle observables whose existence would solely be possible only as a result of the worldsheet foliation structure.  A closely related question, also open for future study, has to do with generalizations to tropological sigma models with worldsheet boundaries, and the classification of topological D-branes in this class of models.  In relativistic topological sigma models, topological D-branes and the corresponding worldsheet boundary conditions were first studied in 1990 in \cite{equivsm}
\footnote{See remarks in Section~3.1 of \cite{ewcsst}.},
in the context of constructing orientifolds of the topological sigma model, not long after the closed-string topological sigma models were first introduced in \cite{ewtsm}.  The basic class of worldsheet boundary conditions describing topological D-branes was shown to involve the choice of a Lagrangian submanifold of the target space.  It should be intriguing to analyze how this construction withstands the process of tropicalization.

Staying within the context of the tropological sigma models constructed in this paper, there are several important open directions to study further.  One is the formalization of the amplitudes in the context of an appropriate generalization of the Atiyah-Segal axioms.  This question actually has two distinct facets:  First, it can be phrased in the context of topological field theories, as studied in Sections~1 to 4 of this paper, and the answer would represent a generalization of Atiyah's axioms \cite{atiyah} for two-dimensional topological field theories.  The worldsheet bordisms should correspond to two-dimensional worldsheets with boundaries, carrying the structure of a foliation; the bordisms should respect this foliation structure, in particular the individual boundary components should be identifiable with individual leaves of the foliation.  The novelty stems from the fact that, unlike in the relativistic case, we have two different kinds of foliation leaves where states can be studied:  The nonsingular leaves of the foliation, and the singular leaves corresponding to the ``location along the sleeve at infinity'', represented by worldsheet punctures.  Perhaps the smoothness property that we discovered around the punctures in \S\ref{sspunct} will be useful in clarifying the question of operator product expansions and factorization when a mixture of nonsingular leaves and punctures is involved in the process.

Alternatively, one can ask this question of axiomatization in the broader context of nonrelativistic conformal theories, as discussed in Section~5.  The answer would then provide a generalization of Segal's axioms of relativistic CFTs \cite{segal} to nonrelativistic CFTs on worldsheets with foliations.  Having such a formalized set of axioms might be useful for extending the list of theories that generalize the construction presented in this paper in ways consistent with the general principles of quantum field theory.
\bigskip

\noindent\textbf{Nonrelativistic limits of relativistic topological sigma models}
\medskip

Since we found that the physical interpretation of taking the tropical limit of the relativistic sigma models corresponds to a certain nonrelativistic version of the relativistic topological sigma models, one may naturally wonder whether our Lagrangian path-integral construction simply follows from taking a direct nonrelativistic limit of the standard relativistic topological theory.  In the particular example of $\MT P^1$ that we focused on in this paper, the answer is yes, although the limit is somewhat subtle -- for example, it requires that the standard auxiliary fields that impose the localization to pseudoholomorphic maps must not be integrated out before the nonrelativistic limit is taken (similarly as in the recently studied case in \cite{ziqi}).  However, we believe that the direct first-principles construction of the tropological path integral that we followed in this paper has several strong advantages over attempting to take a nonrelativistic limit of the relativistic topological sigma model:

(i) In the case of more general target spaces, to find a prescription for taking a nonrelativistic tropical limit of the topological sigma model is not unique, as it requires a choice of preferred coordinates on the target space.  The tropicalization limit then proceeds dimension by dimension, leaving only a discrete group of target-space symmetries (\ref{eeglnz}) that we discussed in \S~\ref{sssym}.  In the process, the general curved metric on the target space becomes piecewise flat, giving the underlying manifold a piecewise linear polyhedral structure in the tropical limit.  Different choices of preferred coordinate systems on the target space will lead to \textit{a priori} different nonrelativistic theories.  

(ii) As we will see in the example presented in \S~\ref{ssgenbj} below, not all tropological sigma models (as defined in this paper) originate from a nonrelativistic limit of a relativistic topological sigma model.  (In the example below, the target space will be of an odd real dimension, and thus not obtainable as a limit of a complex target space.)  An attempt to classify all possible tropological sigma models thus goes clearly beyond the classification of the relativistic sigma models, and it is useful to develop their path integral intrinsically, based on their nonrelativistic symmetries, without relying on nonrelativistic limits of the relativistic theories.

(iii) With the tools developed in this paper, it is also possible to ask what marginal deformations do our worldsheet theories have, and what kind of nontrivial backgrounds in the tropological target space they correspond to.  Again, the classification of possible nontrivial backgrounds in the target space is likely to be wider than the subclass of backgrounds that can be obtained from nonrelativistic limits of backgrounds known in the relativistic case.

\subsection{Generalizations: Targets with more general Jordan structures}
\label{ssgenbj}

Throughout this paper, we have limited our attention to tropological sigma models whose targets are standard tropicalizations of complex manifolds.  However, once we have understood the path integral formulation of such theories and their symmetries, it is natural to ask whether the same formalism can be directly extended to a broader class of target-space geometries, going beyond the idea of tropicalization.  Here we will indicate some such generalizations, without any attempt at completeness.

The Jordan structures $J_i{}^j$ on tropicalized complex manifolds have a special form:  They decompose into two-by-two diagonal blocks, each given in adapted coordinates by (\ref{eejordantar}).  This limitation is non necessary in our path-integral formulation:  We are free to consider target spaces whose Jordan structures contain on the diagonal indecomposable $p\times p$ Jordan blocks with zero eigenvalues,
\be
\begin{pmatrix}0&1&0&\ldots&0&0&0\\
0&0&1&\ldots&0&0&0\\
\vdots&&&\ddots&&&\vdots\\  
0&0&0&\ldots&0&1&0\\
0&0&0&\ldots&0&0&1\\
0&0&0&\ldots&0&0&0\end{pmatrix},
\ee
of arbitrary higher order $p$, not just the lowest-order $p=2$ that appeared in tropicalizations of complex manifolds.

Intriguingly, this means that we are not limited to target spaces whose real dimension is even.  The simplest example $M_3$ of such a target space that goes beyond the limits imposed by tropicalizations of complex manifolds is of real dimension three.  On $M_3$, we will use adapted coordinates $W,X,\Theta$, calling them collectively $Y^i$, with $i=W,X,\Theta$.  In these coordinates, the Jordan structure takes the form
\be
J_i{}^j=\begin{pmatrix}0&1&0\\
0&0&1\\
0&0&0\end{pmatrix}.
\ee
As a result of this higher-order Jordan structure, $M_3$ carries a natural structure of a nested, double foliation:  One-dimensional leaves of constant $X$ and $W$ are embedded into two-dimensional leaves of constant $W$.  

The localization equations for this system are again given by (\ref{eeloceqcov}), which in the adapted coordinates reduces to
\be
\p_\theta X-\p_r W=0,\qquad\p_\theta\Theta-\p_r X=0,\qquad\p_\theta W=0,\qquad\p_\theta X=0.
\ee
They naturally respect the structure of the nested foliation of the target space.  The solutions of the localization equations are locally given by
\bea
W(r,\theta)&=&W_0,\nonumber\\
X(r,\theta)&=&X_0(r),\\
\Theta(r,\theta)&=&\Theta_0(r)+\theta\p_rX_0(r).\nonumber
\eea
Thus, the solutions look just like those for the $(X,\Theta)$ system studied in the bulk of this paper, with each map lying entirely within a two-dimensional leaf of constant $W$.  However, a richer structure emerges when we look at \textit{all} the classical solutions of the bosonic action for the sigma model.

To construct this action, take for simplicity the degenerate metric on $M_3$ such that $G^{XX}=G^{\Theta\Theta}=1$ and zero otherwise.  The rest of the construction then closely parallels that in \S\ref{ssanti}:  First, we introduce the auxiliary fields $\CB^\alpha{}_i$, and write the action using these auxiliaries as
\be
S=\int d^2\sigma \left(\CB^\alpha_iE_\alpha{}^i-\frac{1}{2}\gamma_{\alpha\beta}G^{ij}\CB^\alpha{}_i\CB^\beta{}_j\right).
\ee
Because the number of components of $\CB^\alpha{}_i$ is again redundant, this action exhibits a gauge invariance similar to that generated by $f_+^\alpha{}_i$ in \S\ref{ssanti}.  This gauge symmetry can be simply fixed by setting $\CB^\alpha_W=0$, at the cost of abandoning the full covariance in $Y^i$.  Of the remaining four components of $\CB^\alpha_i$, two enter the action quadratically, and we can integrate them out.  The remaining two enter the action linearly, and we relabel them $\beta_1$ and $\beta_2$, leading to the simplest form of our conformally invariant action
\be
S=\int dr\,d\theta\,\left\{\frac{1}{2}(\p_\theta X-\p_r W)^2+\frac{1}{2}(\p_\theta \Theta-\p_r X)^2+\beta_1\p_\theta W+\beta_2\p_\theta X\right\}.
\ee
The classical equations of motion of this system are locally solved by
\bea
W(r,\theta)&=&W_0(r),\nonumber\\
X(r,\theta)&=&X_0(r),\nonumber\\
\Theta(r,\theta)&=&\Theta_0(r)+\theta\p_r\Theta_1(r),\\
\beta_1(r,\theta)&=&\beta_{1,0}(r)-\theta\p_r{}^2W_0(r),\nonumber\\
\beta_2(r,\theta)&=&\beta_{2,0}(r)+\theta\left(\vphantom{\frac{1}{2}}\p_r\Theta_1(r)-\p_r{}^2X_0(r)\right)\nonumber.
\eea
Note that the general classical solutions are no longer confined to the individual leaf of the codimension-one foliation of fixed $W$, in contrast to the solutions of the localization equations.

\subsection{Generalizations: Coupling to gravity}

In this paper, we focused on the construction of the tropical version of the simplest class of matter systems, the topological sigma models.  In the process, we have uncovered a long list of intriguing differences and similarities of the path integral formulation in this class of two-dimensional field theories, in comparison to the standard relativistic topological sigma models, and its close connections with nonrelativistic quantum field theories on surfaces with nondynamical foliations.  

On the way towards constructing a full-fledged tropicalized version of the path integral for the topological string, the next step would be to couple such matter systems to the appropriate types of worldsheet topological gravity.  Such a coupling is likely to teach us new lessons about the scope of applicability of string theory, perhaps extendable beyond the strict limit of the topological string.  Similarly, on the mathematical side, one of our original motivations came from the work of Mikhalkin, which shows an intriguing isomorphism between the results of the standard evaluation of certain Gromov-Witten invariants in the relativistic setting on one hand, and the tropical analog of this enumerative problem on the other.  The goal has been to shed new light on this result using the worldsheet path-integral methods.  When the result of Mikhalkin is rephrased in the physics language, it is a statement about correlation functions of BRST invariant operators in topological sigma models coupled to topological gravity.  Hence, in order to examine this tropicalization of the Gromov-Witten invariants and Mikhalkin's results from the perspective of the path integral, we also need the matter systems studied in this paper to be coupled to a suitable version of topological gravity.  

Since the tropicalization of the matter sector on the worldsheet is represented by descending from relativistic surfaces with complex structures to the more singular Jordan structures and their associated worldsheet foliations, it is natural to expect that the coupling of such matter systems to topological gravity should be equivalent to making the Jordan structure (and the worldsheet foliation) dynamical.  This will take us to the well-studied realm of nonrelativistic quantum gravities of the Lifshitz type \cite{mqc,lif}, which naturally live on spaces with foliations, albeit without requiring anisotropic worldsheet scaling.  This construction of the tropicalized version of topological gravity and its coupling to tropicalized matter systems brings a host of its own challenges, and we will present the results in the sequel to this paper \cite{sequel}.

\acknowledgments
We wish to thank Ori Ganor and Oleg Viro for illuminating discussions.  E.A. and A.F.V. are thankful to the organizers of the IAS workshop \textit{String Amplitudes at Finite $\alpha'$} (February 2023), Nima Arkani-Hamed, Lorenz Eberhardt and Sebastian Mizera, for the opportunity to attend the workshop, and for their hospitality and illuminating discussions therein.  This work has been supported by NSF grant PHY-2112880.  K.-I.E. also acknowledges support from the NSF Quantum Leap Challenge Institute program through grant OMA-2016245.  

\bibliographystyle{JHEP}
\bibliography{trsm}
\end{document}